\documentclass[amssymb,amsmath,aps,prb,floatfix,tightenlines]{revtex4}
\usepackage{amscd}
\usepackage{dcolumn}
\usepackage[T1]{fontenc}
\usepackage[latin1]{inputenc}
\usepackage{graphicx}
\makeatletter

\renewcommand{\p@subsection}{}

\renewcommand{\p@subsubsection}{}

\makeatother
\def\be{\begin{equation}}
\def\ee{\end{equation}}
\def\ba{\begin{eqnarray}}
\def\ea{\end{eqnarray}}
\def\>{\rangle}
\def\<{\langle}
\def\n{\nonumber}
\def\lb{\left[}
\def\rb{\right]}
\def\<{\langle}
\def\>{\rangle}

\newcommand{\id}{\openone}
\begin{document}
\title{Entanglement in the interaction between two quantum oscillator systems}
\author{
        ILki~Kim and\, Gerald~J.~Iafrate}
\affiliation{\vspace*{0.ex}\hspace*{-.7cm}Department of
             Electrical and Computer Engineering,
             North Carolina State University, Raleigh,
             NC 27695}
%
%
\begin{abstract}
The fundamental quantum dynamics of two interacting oscillator
systems are studied in two different scenarios. In one case, both
oscillators are assumed to be linear, whereas in the second case,
one oscillator is linear and the other is a non-linear,
angular-momentum oscillator; the second case is, of course, more
complex in terms of energy transfer and dynamics. These two
scenarios have been the subject of much interest over the years,
especially in developing an understanding of modern concepts in
quantum optics and quantum electronics. In this work, however,
these two scenarios are utilized to consider and discuss the
salient features of quantum behaviors resulting from the
interactive nature of the two oscillators, i.e., coherence,
entanglement, spontaneous emission, etc., and to apply a {\em
measure of entanglement} in analyzing the nature of the
interacting systems.

The Heisenberg equation for both coupled oscillator scenarios are
developed in terms of the relevant reduced kinematics operator
variables and parameterized commutator relations. For the second
scenario, by setting the relevant commutator relations to one or
zero, respectively, the Heisenberg equations are able to describe
the full quantum or classical motion of the interaction system,
thus allowing us to discern the differences between the fully
quantum and fully classical dynamical picture.

For the coupled linear and angular-momentum oscillator system in
the fully quantum-mechanical description, we consider special
examples of two, three, four-level angular momentum systems,
demonstrating the explicit appearances of entanglement. We also
show that this entanglement persists even as the {\em coupled}
angular momentum oscillator is taken to the limit of a large
number of levels, a limit which would go over to the classical
picture for an {\em uncoupled} angular momentum oscillator.
\end{abstract}
\maketitle
\section{Introduction}
In this paper, we study the quantum dynamics of two interacting
oscillator systems in two different, but very well-known
scenarios. In case one, both oscillators are assumed to be linear,
and in case two, one oscillator is assumed to be linear, while the
other oscillator is assumed to be non-linear (angular momentum
oscillator). We study Heisenberg equations and the time evolution
of the wave-functions $\left|\psi(t)\right\> =
e^{-i\,\hat{H}\,t/\hbar}\,\left|\psi(0)\right\>$ for both cases to
discuss the quantum interactive behaviors between the two
oscillators therein, e.g., entanglement, spontaneous emission (in
case two), and quantum versus classical differences (in case two).
Entanglement, an implicit non-classical ingredient of interacting
quantum systems, is the key to enabling quantum computation, which
allows for the solution of certain classes of problems more
efficiently than the classical counterpart (see, e.g.,
Refs.~\onlinecite{SHO94,GRO96}). For case two, by setting the
relevant commutator relation in each oscillator to one or zero, we
utilize an interesting quantum-classical scheme \cite{SEN71} to
clearly characterize the quantum features in the system.

In order to investigate the time evolution of the wave-function
and the notion of entanglement in case one, we adopt the
coupled-boson representation,\cite{SCW01} which was discussed by
Schwinger in terms of the correspondence between two
one-dimensional linear oscillators and an angular momentum
oscillator. Based on this representation, we will easily obtain
exact energy eigenvalues and eigenstates of case one. The
coupled-boson representation has been used in context with similar
types of problems to construct an ideal model for
quantum-mechanical interference\cite{SHA73} and to study the
current operator in two coupled linear oscillator (the rate of
exchange of occupation number between oscillators).\cite{IAF75}

We adopt here $1 - \mbox{Tr}\,(\hat{\rho}^{(\nu)})^2$ as an
entanglement measure for the total system in a pure
state,\cite{RUD01} where $\hat{\rho}^{(\nu)}$ is a reduced density
matrix of individual oscillator $\nu$. We will show that this
measure increases with the occupation number $n$ for a given
initial state for case one, and with the maximum quantum number
$j$ of the angular momentum oscillator for case two, respectively;
this increase with occupation number results from considering a
dissipationless and coherent process, whereas we know that
entanglement will diminish when loss is taken into account, e.g.,
resulting from the (thermal) interaction between the system and
its environment.\cite{GIU96} For case two, we will discuss
entanglement for the cases where the angular momentum oscillator
is in the states $j=\frac{1}{2},\,1,\,\frac{3}{2}$ explicitly.

In section \ref{sec:linear_linear_osc}, we study the interaction
between two linear oscillators. The Heisenberg equations of motion
are analyzed in the coupled boson representation, leading to the
straightforward diagonalization of the coupled oscillator
Hamiltonian; from this diagonalized Hamiltonian, the exact
eigenstates and eigenvalues of the coupled oscillator Hamiltonian
are obtained, thus allowing for the calculation of the reduced
density matrix, and the subsequent study of the entanglement under
the two initial conditions of simple product state (a disentangled
state) and exact eigenstate for the coupled oscillator system (an
entangled state).

In section \ref{sec:linear_non_linear}, we study the interaction
between a linear and a non-linear, angular momentum oscillator. We
make use of the seminal work of Ref. \onlinecite{SEN71} to analyze
the quantum-mechanical and the classical dynamics of this
interaction, and to connect the quantum manifestation of
spontaneous emission in this system with the notion of
entanglement. Further, the total system Hamiltonian is
diagonalized for $j=\frac{1}{2}, 1, \frac{3}{2}$ using rotational
generating operators of the group $SU(n)$.\cite{MAH98} From this
diagonalization, the reduced density matrix, and thus the measure
of entanglement, is calculated for simple product (disentangled)
initial states. Unlike the linear-linear oscillator system, this
coupled system displays periodicity in the measure of entanglement
for $j=\frac{1}{2}$ and $j=1$, whereas for $j=\frac{3}{2}$ and
beyond, the measure of entanglement is aperiodic; this
aperiodicity is shown to be a result of the complex structure of
the diagonalized multi-level Hamiltonian.

%
\section{Interaction of two linear oscillators}\label{sec:linear_linear_osc}
\subsection{Quantum behavior in the Heisenberg picture}
Let us begin with case one. The coupled oscillator system under
investigation is described in the {\em rotating wave
approximation} by the Hamiltonian
\begin{eqnarray}\label{eq:hamiltonian1}
    \hat{H}\; =\; {\textstyle \hbar\,\omega_1
    \left(\hat{a}_1^{\dagger}\,\hat{a}_1\,
    +\, \frac{1}{2} \id\right)\, +\, \hbar\,\omega_2
    \left(\hat{a}_2^{\dagger}\,\hat{a}_2\, +\, \frac{1}{2}
    \id\right)\, +\, \hbar\,\kappa
    \left(\hat{a}_1^{\dagger}\,\hat{a}_2\,
    +\, \hat{a}_1\,\hat{a}_2^{\dagger}\right)\,,}
\end{eqnarray}
where $\kappa$ denotes a coupling strength, and we assume that
$\omega_1 \geq \omega_2$. Schwinger found that the
quantum-mechanical angular momentum can be obtained by using
creation and annihilation operators of two one-dimensional
harmonic oscillators in terms of $\hat{\mathcal{J}}_{l} \propto
\hat{a}^{\dagger} \otimes \hat{a}$. This is explicitly given as
follows:\cite{SCW01}
\begin{equation}\label{eq:bosonic_rep}
    {\textstyle \hat{\mathcal{J}}_x\, \equiv\, \frac{\hbar}{2}
    \left(\hat{a}_1^{\dagger} \otimes \hat{a}_2 +
    \hat{a}_1 \otimes \hat{a}_2^{\dagger}\right)}\,,\;
    {\textstyle \hat{\mathcal{J}}_y\, \equiv\, \frac{\hbar}{2i}
    \left(\hat{a}_1^{\dagger} \otimes \hat{a}_2 -
    \hat{a}_1 \otimes \hat{a}_2^{\dagger}\right)}\,,\;
    {\textstyle \hat{\mathcal{J}}_z\, \equiv\, \frac{\hbar}{2} \left(\hat{n}_1 -
    \hat{n}_2\right)}\,,\;
    {\textstyle \hat{\mathcal{J}}\, \equiv\, \frac{\hbar}{2} \left(\hat{n}_1 +
    \hat{n}_2\right)\,,}
\end{equation}
where $\hat{n}_{\nu} = \hat{a}_{\nu}^{\dagger}\,\hat{a}_{\nu}$
with $\nu=1,2$ is obviously an occupation number operator of each
linear oscillator $\nu$. From $\lb \hat{a}_{\mu}\,,
\hat{a}_{\nu}^{\dagger} \rb = \id\,\delta_{\mu\nu}$ the operators
$\hat{\mathcal{J}}_{l}$, where $l=x,y,z$, satisfy the usual
angular-momentum commutation relations. The operator
$\hat{\mathcal{J}}$ yields the quantum number of angular momentum,
$j=0,\frac{1}{2},1,\frac{3}{2}, \cdots$. In terms of the angular
momentum operators in (\ref{eq:bosonic_rep}) the Hamiltonian in
(\ref{eq:hamiltonian1}) is rewritten as
\begin{equation}\label{eq:coupled_boson_hamiltonian}
    \hspace*{-3mm}{\textstyle \hat{H}\; =\; \Delta \omega\,\hat{\mathcal{J}}_z\, +\,
    2\,\kappa\,\hat{\mathcal{J}}_x\,
    +\, \left(\omega_1+\omega_2\right)\,\hat{\mathcal{J}}\, +\,
    \frac{\id}{2}\,\hbar\,\left(\omega_1 + \omega_2\right)\,,}
\end{equation}
where $\Delta \omega \equiv \omega_1-\omega_2 \geq 0$. Since
$[\hat{\mathcal{J}}_{l},\hat{\mathcal{J}}\,] = 0$, we easily
obtain a constant of motion $\hat{\mathcal{J}}$ with $j =
\frac{1}{2}(n_1+n_2)$\,:
\begin{equation}\label{eq:equation_of_motion1}
    {\textstyle i\,\hbar\,\dot{\hat{\mathcal{J}}}\; =\;
    \left[\hat{\mathcal{J}}\,,\,\hat{H}\right]\; =\; 0\,,}
\end{equation}
which means that the total oscillator occupation number is
conserved. Also, we find that ${\textstyle
\dot{\hat{\mathcal{J}}}_x = - \Delta \omega\,\hat{\mathcal{J}}_y}$
and so $\hat{\mathcal{J}}_x$ is a constant of motion for $\Delta
\omega = 0$.

With the aid of the identity,
\begin{equation}\label{eq:identity_1}
    {\textstyle e^{\vartheta \hat{U}}\, \hat{V}\, e^{-\vartheta \hat{U}}}\;
    =\; {\textstyle \hat{V}\,
    +\, \vartheta [\hat{U},\,\hat{V}]\, +\,
    \frac{\vartheta^2}{2!}\,
    [\hat{U},\,[\hat{U},\,\hat{V}]]\, +\, \frac{\vartheta^3}{3!}\,
    [\hat{U},\,[\hat{U},\,[\hat{U},\,\hat{V}]]]\, +\,
    \cdots\,,}
\end{equation}
we evaluate $e^{i \gamma \hat{\mathcal{J}}_y(0)/\hbar}\;
{\hat{H}}\; e^{-i \gamma \hat{\mathcal{J}}_y(0)/\hbar}$, where $0
\leq \gamma \leq \frac{\pi}{2}$, to arrive at the diagonalized
Hamiltonian of eq. (\ref{eq:coupled_boson_hamiltonian}) as
\begin{equation}\label{eq:diagonal_hamiltonian}
    \hspace*{-3mm}{\hat{H}}_d\; =\; {\textstyle e^{i \gamma \hat{\mathcal{J}}_y(0)/\hbar}\;
    {\hat{H}}\; e^{-i \gamma \hat{\mathcal{J}}_y(0)/\hbar}\; =\;
    \frac{\id}{2}\,\hbar\,\left(\omega_1 + \omega_2\right)\, +\,
    \left\{\Delta \omega \cdot \left(\cos \gamma\right)\, +\,
    2 \kappa \cdot \left(\sin
    \gamma\right)\right\}\,\hat{\mathcal{J}}_z(0)\,
    +\, \left(\omega_1\, +\, \omega_2\right)\,\hat{\mathcal{J}}\,,}
\end{equation}
where $\tan \gamma \equiv \frac{2 \kappa}{\Delta \omega}$ (note
that in case of $\Delta \omega = 0$, we have $\gamma =
\frac{\pi}{2}$ for any non-zero $\kappa$). From eqs.
(\ref{eq:bosonic_rep}), (\ref{eq:diagonal_hamiltonian}) we now
obtain two renormalized uncoupled harmonic oscillators with the
renormalized energy splitting for each oscillator,
\begin{equation}\label{eq:uncoupled_hamiltonian}
    {\textstyle {\hat{H}}_d\; =\; \hbar\,\omega_1'\,\left(\hat{A}_1^{\dagger}\,\hat{A}_1\,
    +\, \frac{\id}{2}\right)\, +\, \hbar\,\omega_2'\,\left(\hat{A}_2^{\dagger}\,\hat{A}_2\,
    +\, \frac{\id}{2}\right)}
\end{equation}
where $\omega_1' \equiv \frac{1}{2}(\omega_1+\omega_2) +
\frac{1}{2}\bar{\omega}$\,, $\omega_2' \equiv
\frac{1}{2}(\omega_1+\omega_2) - \frac{1}{2}\bar{\omega}$. Here,
$\bar{\omega}^2 = \left(2 \kappa\right)^2 + (\Delta \omega)^2$,
and $\hat{A}_{\nu}(t)\, \equiv\, e^{-i \gamma
\hat{\mathcal{J}}_y(0)/\hbar}\; \hat{a}_{\nu}(t)\; e^{i \gamma
\hat{\mathcal{J}}_y(0)/\hbar}$ explicitly reads
\begin{equation}\label{eq:rotation_mat}
    {\textstyle \left( \begin{array}{c}
                           \hat{A}_{1}(t)\\
                           \hat{A}_{2}(t)
                       \end{array} \right)\; =\;
    \hat{R}\left(\frac{-\gamma}{2}\right) \cdot
    \left( \begin{array}{c}
               \hat{a}_{1}(t)\\
               \hat{a}_{2}(t)
           \end{array} \right)}
\end{equation}
where the rotation matrix is
\begin{equation}
    {\textstyle \hat{R}(\vartheta)\; =\;
    \left( \begin{array}{lr}
               \cos \vartheta  &  -\sin \vartheta\\
               \sin \vartheta  &  \cos \vartheta
           \end{array} \right)\,.}
\end{equation}
Thus, from the Schwinger approach, we realize that the coupling
term in eq. (\ref{eq:hamiltonian1}) can be equivalently
represented as the $\hat{\mathcal{J}}_x$ component of angular
momentum, and therefore, $\hat{H}$ in eq. (\ref{eq:hamiltonian1})
can be diagonalized by a unitary rotational operator with respect
to the angular momentum operator~$\hat{\mathcal{J}}_y(0)$ through
$\gamma$.\cite{IAF75} Then, using eqs.
(\ref{eq:uncoupled_hamiltonian}), (\ref{eq:rotation_mat}) we
easily obtain the original annihilation operators as
\begin{equation}\label{eq:a_and_a_dagger}
    \hat{a}_{\nu}(t)\; =\;
    {\textstyle \left\{\left(\cos \frac{\gamma}{2}\right)^2\,
    e^{-i \omega_{\nu}'\,t}\, +\, \left(\sin \frac{\gamma}{2}\right)^2\,
    e^{-i \omega_{\mu}'\,t}\right\}\, \hat{a}_{\nu}(0)\,
    +\, \frac{1}{2}\,\left(\sin \gamma\right)\,\left(e^{-i \omega_1'\,t}\,
    -\, e^{-i \omega_2'\,t}\right)\, \hat{a}_{\mu}(0)\,,}
\end{equation}
where $\nu, \mu = 1,2$ but $\nu \ne \mu$. Here, neglecting the
``non-local'' observable $\hat{a}_{\mu}(0)$, we obviously have
$[\hat{a}_{\nu}(t), \hat{a}_{\nu}^{\dagger}(t)] \ne 1$ at $t \ne
0$. Now, we easily acquire the exact eigenstates and eigenvalues
of the Hamiltonian in (\ref{eq:hamiltonian1}) as
\begin{equation}\label{eq:eigenstates}
    {\textstyle \left|n_1,n_2\right\>_{H}\; \equiv\; e^{-i \gamma
    \hat{\mathcal{J}}_y(0)/\hbar}\,
    \left|n_1\right>\left|n_2\right\>}\;;\;\;
    {\textstyle E_{n_1,n_2}\; =\; \frac{\hbar}{2}\,(n_1 + n_2 + 1)\,(\omega_1 +
    \omega_2)\, +\, \frac{\hbar}{2}\,(n_1 - n_2)\,\bar{\omega}\,.}
\end{equation}
We see here that for $\gamma = \frac{\pi}{2}$ (i.e., $\Delta
\omega = 0$), the eigenstates $\left|n_1,n_2\right\>_{H}$ will
also be eigenstates of $\hat{\mathcal{J}}_x(0) =
\hat{\mathcal{J}}_x(t)$.

\subsection{Appearance of the entanglement in the Schr\"{o}dinger picture}
Let us now study the entanglement in the time evolution of the two
coupled linear oscillators, $|\psi(t)\> = e^{-i \hat{H}
t/\hbar}\,|\psi(0)\>$. To this end, we adopt,\cite{RUD01} as a
measure of entanglement,
\begin{equation}\label{eq:purity_measure}
    {\textstyle \mathcal{M}_{|\psi(t)\>}\; =\; 1\, -\,
    \mathcal{P}\left[\hat{\rho}^{(\nu)}(t)\right]\,,}
\end{equation}
where $\mathcal{P}\left[\hat{\rho}^{(\nu)}\right] \equiv
\mbox{Tr}_{\nu}\, (\hat{\rho}^{(\nu)})^2$ is the purity measure of
the state of each oscillator $\nu$. Here, it clearly holds that
$\mathcal{P}[\hat{\rho}^{(1)}] =
\mathcal{P}[\hat{\rho}^{(2)}]$.\cite{HUG93} As can also easily be
verified, $\mathcal{P}[\hat{\rho}^{(\nu)}]$ takes on a maximum
value of $1$ if the oscillator $\nu$ is in a pure state; it then
follows that $\mathcal{M}_{|\psi\>} = 0$ for a simple product
(disentangled) state, ${\textstyle \left|\psi\right\> =
|\psi^{(1)}\> |\phi^{(2)}\>}$. On the other hand,
$\mathcal{P}[\hat{\rho}^{(\nu)}]$ takes on its minimum value
$\frac{1}{M}$, where $M$ is a number of accessible orthogonal
states $\{|p_{\nu}\>; p=1,2,\cdots, M\}$ of each oscillator $\nu$,
if the oscillator $\nu$ is in the maximally mixed state given by
$\hat{\rho}_{kl}^{(\nu)} = \frac{1}{M}\,\delta_{kl}$; then, the
corresponding total wave-function $\left|\psi\right\>$ is a
maximally entangled state in form of
$\frac{1}{\sqrt{M}}\,(|1_{1}\>|1_{2}\> + |2_{1}\>|2_{2}\> + \cdots
+ |M_{1}\>|M_{2}\>)$, and $\mathcal{M}_{|\psi\>}$ has its maximum
value $\frac{M-1}{M}$. We will explicitly determine
$\mathcal{M}_{|\psi(t)\>}$ in its time evolution with
mathematically manageable initial states $|\psi(0)\>$ given below.

We obtain the reduced density matrix, $\hat{\rho}^{(\nu)}(t)$,
from the system state evolution, $|\psi(t)\>$, by constructing the
system density matrix
$\hat{\rho}(t)=\left|\psi(t)\right>\left<\psi(t)\right|$ from the
known state of the system, and then taking the trace of
$\hat{\rho}(t)$ with respect to all subsystem matrix elements
within the system but those of subsystem $\nu$. For our oscillator
system under study, the state evolution is given by
\begin{equation}\label{eq:schroedinger_pic_psi1}
    \left|\psi(t)\right>\; =\; \sum_{n_1,n_2=0}^{\infty}
    c_{n_1,n_2}\; \left|n_1,n_2\right>_{H}\;
    e^{-\frac{i}{\hbar}E_{n_1,n_2}\,t}\,,
\end{equation}
where $c_{n_1,n_2} =
\left\<\psi(0)\right|\left.n_1,n_2\right\>_{H}^{\ast}$. Therefore,
in constructing $\hat{\rho} = \left|\psi\right>\left<\psi\right|$
from (\ref{eq:schroedinger_pic_psi1}), and taking the trace of
$\hat{\rho}$ with respect to oscillator $2$, we formally acquire
the reduced density matrix for oscillator $1$ as
\begin{equation}\label{eq:schroedinger_pic_psi2}
    \hat{\rho}_{kl}^{(1)}(t)\; =\; \mbox{Tr}_2\, \hat{\rho}(t)\;
    =\; \sum_{s=0}^{\infty}\, \left\{\hat{\rho}(t)\right\}_{ks;ls}\,,
\end{equation}
where $\left\{\hat{\rho}(t)\right\}_{ks;ls} \equiv
\left<s_{2}\right|\left<k_{1}\right|\hat{\rho}(t)\left|l_{1}\right>\left|s_{2}\right>$.
Using $|\psi(t)\>$ in eq. (\ref{eq:schroedinger_pic_psi1}), we
explicitly obtain
\begin{equation}\label{eq:reduced_density_mat1}
    \hspace*{-7.mm}\displaystyle{\hat{\rho}_{kl}^{(1)}(t)\; =\; \sum_{n_1,n_2}\,
    \sum_{n_1',n_2'}\, c_{n_1,n_2} \cdot c_{n_1',n_2'}^{\ast} \cdot
    e^{-\frac{i}{\hbar}(E_{n_1,n_2}-E_{n_1',n_2'})\,t}\,
    \cdot\, \sum_{s}\,
    \left<s_{2}\right|\left<k_{1}\right|\,e^{-i \gamma
    \hat{\mathcal{J}}_y/\hbar}\,\left|n_1\right>\left|n_2\right> \cdot
    \left<n_2'\right|\left<n_1'\right|\,e^{i \gamma
    \hat{\mathcal{J}}_y/\hbar}\,\left|l_{1}\right>\left|s_{2}\right>}\,,
\end{equation}
where $\hat{\mathcal{J}}_y = \hat{\mathcal{J}}_y(0)$, and
$|n_1,n_2\>_{H}$ from eq. (\ref{eq:eigenstates}) has been used in
eq. (\ref{eq:schroedinger_pic_psi1}) to obtain the explicit
result.

Now, in transforming eq. (\ref{eq:reduced_density_mat1}) to the
coupled-boson representation discussed in eq.
(\ref{eq:bosonic_rep}), we relabel $|n_1\>|n_2\>$ and
$|n_1'\>|n_2'\>$ in terms of the eigenstates of
$\hat{\mathcal{J}}_z$, namely, $|j;m\>$ and $|j';m'\>$,
respectively, where
$j=\frac{1}{2}(n_1+n_2),\,m=\frac{1}{2}(n_1-n_2)$ and
$j'=\frac{1}{2}(n_1'+n_2'),\,m'=\frac{1}{2}(n_1'-n_2')$. Since the
rotation with respect to $\hat{\mathcal{J}}_y$ conserves the
quantum number $j$ as noted in eq. (\ref{eq:equation_of_motion1}),
we also find that
\begin{equation}\label{eq:constraints}
    k\, +\, s\; =\; 2\,j\,;\; l\, +\, s\; =\; 2\,j'\,.
\end{equation}
Taking into account that $m = -j, -j+1, \cdots, j$ and $m' = -j',
-j'+1, \cdots, j'$ for a given $j,\,j'$, respectively, it follows
that eq.~(\ref{eq:reduced_density_mat1}) can be reexpressed as
\vspace*{-1.5mm}
\begin{equation}\label{eq:reduced_density_mat2}
    \hspace*{-12mm}\hat{\rho}_{kl}^{(1)}(t)\; =\; \sum_{j,j'=0}^{\infty}\;
    \sum_{m=-j}^{j}\; \sum_{m'=-j'}^{j'}\, c_{j,m}\; \cdot\;
    c_{j',m'}^{\ast}\; \cdot\;
    e^{-\frac{i}{\hbar}(E_{j+m,j-m}-E_{j'+m',j'-m'})\,t}\; \cdot\;
    \left<j;k-j\right|e^{-i \gamma \hat{\mathcal{J}}_y/\hbar}\left|j;m\right>
    \left<j';m'\right|e^{i \gamma \hat{\mathcal{J}}_y/\hbar}\left|j';l-j'\right>
\end{equation}
where $k = 0, 1, \cdots, 2j$ and $l = 0, 1, \cdots, 2j'$,
respectively. Then, in defining $\mathbf{k} \equiv k-j$ and
$\mathbf{l} \equiv l-j$ we relabel $\hat{\rho}_{kl}^{(1)}$ in
terms of $\hat{\rho}_{\mathbf{k}\mathbf{l}}^{(1)}$, where
$\mathbf{k} = -j, -j+1, \cdots, j$ and $\mathbf{l} = -j', -j'+1,
\cdots, j'$, respectively. This clearly shows that two coupled
harmonic oscillators can naturally be treated as an angular
momentum oscillator. We now make use of the coupled-boson algebra
to evaluate the reduced density matrix
$\hat{\rho}_{\mathbf{k}\mathbf{l}}^{(1)}(t)$ below for two
physically interesting, transparent, and mathematically manageable
initial states, $|\psi(0)\>$, to illustrate the entanglement
measure $\mathcal{M}_{|\psi(t)\>}$, namely, the case of a simple
product state, $\left|N_1\right>\left|N_2\right>$ (case I), and
the case of an eigenstate of the exact Hamiltonian in eq.
(\ref{eq:coupled_boson_hamiltonian}), $\left|N_1,N_2\right>_{H}$
of eq. (\ref{eq:eigenstates}) (case II). In terms of physical
interest, the simple product state in case I is an initial
condition which inherently assumes no entanglement at time zero,
whereas case II, using the exact eigenstate of the Hamiltonian,
implicitly includes entanglement.

\begin{figure}[htbp]
    {\par\centering \resizebox*{0.45\textwidth}{!}
    {\includegraphics[angle=-90]{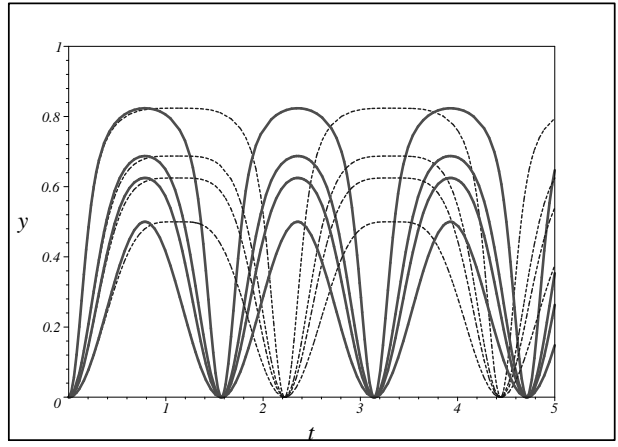}} \par}
    \caption{\label{fig:reduced_mat_time_evolution}
    The measure of entanglement, $y(t) = \mathcal{M}_{|\psi(t)\>}$ given in
    eq. (\ref{eq:entanglement_measure_case_1}) versus time
    $t$ is depicted for $\kappa = 1$ and initial states $|\psi(0)\> = |N\>|0\> =
    |J;J\>$, where
    $N = 1, 2, 3, 10$ (i.e., $J = \frac{1}{2}, 1, \frac{3}{2}, 5$)
    in progressive order
    from bottom to top; the solid lines are obtained for $\gamma =
    \frac{\pi}{2}$ (or $\omega_1 = \omega_2$), while the dash lines for $\gamma =
    \frac{\pi}{4}$ (or $\omega_1-\omega_2 = 2$).}
\end{figure}

Let us begin with the case I: $|\psi(0)\> =
\left|N_1\right>\left|N_2\right>$. In the coupled-boson
representation, this state becomes $\left|J;M\right>$ where
$J=\frac{1}{2}(N_1+N_2)$ and $M=\frac{1}{2}(N_1-N_2)$. We now have
$c_{j,m} = \left<J;M\right|\left.j;m\right>_H^{\ast} \cdot
\delta_{jJ}$ and $c_{j',m'}^{\ast} =
\left<J;M\right|\left.j';m'\right>_{H} \cdot \delta_{j'J}$ in eq.
(\ref{eq:reduced_density_mat2}) which then reduces to the diagonal
form,
\begin{equation}\label{eq:reduced_density_mat3}
    \hspace*{-7mm}\hat{\rho}_{\mathbf{k}\mathbf{l}}^{(1)}(t)\; =\;
    \delta_{\mathbf{k}\mathbf{l}} \sum_{m,m'=-J}^{J}
    \left<J;M\right|\left.J;m\right>_H^{\ast}\,
    \left<J;M\right|\left.J;m'\right>_{H}\,
    {\textstyle \times\;
    \left<J;\mathbf{k}\right|\left.J;m\right>_{H}\,
    \left<J;\mathbf{l}\right|\left.J;m'\right>}_H^{\ast}\;
    e^{i (m-m') \bar{\omega} t}\,,
\end{equation}
where $\delta_{\mathbf{k}\mathbf{l}}$ results from
eq.~(\ref{eq:constraints}) with $j=j'=J$. In eq.
(\ref{eq:reduced_density_mat3}), all the matrix elements can be
written in terms of $\left<J;\mu_2\right|\left.J;\mu_1\right>_H
\equiv U_{\mu_1\,\mu_2}^{(J)}(\gamma)$ with ($\mu_1 = m, m'$) and
($\mu_2 = M, \mathbf{k}, \mathbf{l}$), where
$U_{\mu_1\,\mu_2}^{(J)}(\gamma)$ is formally explicitly expressed
as \cite{SCW01}
\begin{equation}\label{eq:rotation_mat1}
    \hspace*{-5mm}{\textstyle{U_{\mu_1\,\mu_2}^{(J)}(\gamma)\, =\,
    \sqrt{\frac{(J+\mu_2)!}{(J-\mu_2)!}}}}\,
    \frac{(\sin \frac{\gamma}{2})^{\mu_1-\mu_2}\,
    (\cos \frac{\gamma}{2})^{-\mu_1-\mu_2}}{(2^{J+\mu_2})\,
    \sqrt{(J+\mu_1)!\, (J-\mu_1)!}}
    {\textstyle\; \times\; \left(\frac{d}{d \cos \gamma}\right)^{J-\mu_2}
    \left\{(\cos \gamma + 1)^{J+\mu_1}\, (\cos \gamma -
    1)^{J-\mu_1}\right\}}\,;
\end{equation}
here we see that $U_{\mu_2\,\mu_1}^{(J)}(\gamma) =
U_{\mu_1\,\mu_2}^{(J)}(-\gamma)$, and
$\{U_{\mu_1\,\mu_2}^{(J)}(\gamma)\}^{\ast} =
U_{\mu_1\,\mu_2}^{(J)}(\gamma)$. Particularly for the case that
$\mu_2=J$, this reduces further to
\begin{equation}\label{eq:rotation_mat2}
    \hspace*{-3mm}\textstyle U_{\mu J}^{(J)}(\gamma)\, =\, {\textstyle \sqrt{\frac{(2J)!}{(J+\mu)!\, (J-\mu)!}}\,
    \left(-\sin \frac{\gamma}{2}\right)^{J-\mu}\,
    \left(\cos \frac{\gamma}{2}\right)^{J+\mu}\,,}
\end{equation}
which will be used later. Equivalently,
$U_{\mu_1\,\mu_2}^{(J)}(\gamma)$ can be expressed in terms of the
tabulated Jacobi polynomial \cite{ABR64}
$\mathcal{P}_{n}^{(\alpha,\beta)}(x)$ in the form
\begin{equation}\label{eq:rotation_mat3}
    \hspace*{-5mm}\textstyle{U_{\mu_1\,\mu_2}^{(J)}(\gamma)}\; =\;
    \textstyle{(-1)^{\mu_1+\mu_2}\,
    \sqrt{\frac{(J+\mu_2)!\, (J-\mu_2)!}{(J+\mu_1)!\, (J-\mu_1)!}}\,
    \left(\sin \frac{\gamma}{2}\right)^{\mu_2-\mu_1}\, \times}\;
    \textstyle{\left(\cos \frac{\gamma}{2}\right)^{\mu_1+\mu_2}
    \cdot \left\{{\mathcal{P}}_{J-\mu_2}^{(\mu_2-\mu_1,\mu_1+\mu_2)}(\cos
\gamma)\right\}}\,.
\end{equation}
Reexpressing $\hat{\rho}_{\mathbf{k}\mathbf{l}}^{(1)}$ in eq.
(\ref{eq:reduced_density_mat3}) in terms of
$U_{\mu_1\,\mu_2}^{(J)}(\gamma)$, we obtain
\begin{equation}\label{eq:reduced_density_mat3_1}
    \hspace*{-15mm}\hat{\rho}_{\mathbf{k}\mathbf{l}}^{(1)}(t)\; =\;
    \delta_{\mathbf{k}\mathbf{l}} \sum_{m,m'=-J}^{J}\,
    \{U_{mM}^{(J)}(\gamma)\}^{\ast}\, \cdot\,
    U_{m'M}^{(J)}(\gamma)\, \cdot\,
    U_{m\mathbf{k}}^{(J)}(\gamma)\, \cdot\,
    \{U_{m'\mathbf{l}}^{(J)}(\gamma)\}^{\ast}\, \cdot\,
    e^{i (m-m') \bar{\omega} t}\; =\; {\textstyle \delta_{\mathbf{k}\mathbf{l}}\,
    \cdot\, f_{\mathbf{l}}^{(1)}(t)\, \cdot\, \{f_{\mathbf{k}}^{(1)}(t)\}^{\ast}}\; \equiv\;
    {\textstyle \delta_{\mathbf{k}\mathbf{l}}\, \cdot\,
    |f_{\mathbf{k}}^{(1)}(t)|^2}\,,
\end{equation}
where
\begin{equation}\label{eq:f_function}
    f_{\mathbf{k}}^{(1)}(t)\, \equiv\, \sum_{m=-J}^{J}\,
    U_{mM}^{(J)}(\gamma) \cdot
    U_{m\mathbf{k}}^{(J)}(\gamma) \cdot e^{-i m \bar{\omega} t}\\
\end{equation}
with the property that $f_{\mathbf{k}}^{(1)}(0) =
\delta_{M\mathbf{k}}$ [here, we used the relation that
$\{U_{mM}^{(J)}(\gamma)\}^{\ast} = U_{mM}^{(J)}(\gamma)$];
$f_{\mathbf{k}}^{(1)}(t)$ can be explicitly evaluated for any set
of the harmonic oscillator initial condition numbers,
$\{N_1,\,N_2\}$. We refer the reader to the Appendix
\ref{sec:appendix1} for the explicit expression of eq.
(\ref{eq:f_function}) obtained from the substitution of eq.
(\ref{eq:rotation_mat3}), which will be useful below. From
(\ref{eq:reduced_density_mat3_1}) we obtain the measure of
entanglement given by
\begin{equation}\label{eq:entanglement_measure_case_1}
    {\textstyle \mathcal{M}_{|\psi(t)\>}\; = \; 1\, -\,} \sum_{\mathbf{k}=-J}^{J}\,
    {\textstyle |f_{\mathbf{k}}^{(1)}(t)|^4\, \geq\, 0\,.}
\end{equation}
We note generally that $f_{\mathbf{k}}^{(1)}$ in eq.
(\ref{eq:f_function}) is a time-dependent oscillatory function
with a modal distribution coefficient $U_{mM}^{(J)}(\gamma) \cdot
U_{m\mathbf{k}}^{(J)}(\gamma)$ governing the oscillatory strength.
This clearly influences the measure of entanglement as noted in
eq. (\ref{eq:entanglement_measure_case_1}) and as seen in Fig.
\ref{fig:reduced_mat_time_evolution} for specific values of $J$.

Let us now apply eqs. (\ref{eq:reduced_density_mat3_1}) and
(\ref{eq:entanglement_measure_case_1}) for specific values of $J$.
For $J = \frac{1}{2}$, the total occupation number $N_1+N_2 = 1$,
there are two possible simple product states
$\left|\frac{1}{2};\frac{1}{2}\right>,
\left|\frac{1}{2};-\frac{1}{2}\right>$ describing this initial
condition. We can easily evaluate $\hat{\rho}^{(1)}(t)$: for
$\left|\psi_1(0)\right> = \left|\frac{1}{2};\frac{1}{2}\right>$,
from eqs. (\ref{eq:rotation_mat2}) and (\ref{eq:f_function}) we
obtain $f_{-\frac{1}{2}}^{(1)}(t) = \cos \frac{\bar{\omega} t}{2}
+ i\,(\cos \gamma)\,\sin \frac{\bar{\omega} t}{2}$, and then the
$2 \times 2$ diagonal matrix from eq.
(\ref{eq:reduced_density_mat3_1}),
\begin{equation}\label{eq:reduced_density_mat4}
    \hspace*{-5mm}{\textstyle \hat{\rho}_{-\frac{1}{2},-\frac{1}{2}}^{(1)}\, =\,
    \frac{1}{2} \left(\sin \gamma\right)^2\,
    \left\{1-\cos\left(t \bar{\omega}\right)\right\}\,;\;
    \hat{\rho}_{\frac{1}{2},\frac{1}{2}}^{(1)}\, =\, 1\, -\,
    \hat{\rho}_{-\frac{1}{2},-\frac{1}{2}}^{(1)}\,.}
\end{equation}
Similarly, for $\left|\psi_2(0)\right> =
\left|\frac{1}{2};-\frac{1}{2}\right>$, we acquire the diagonal
matrix $\hat{\rho}_2^{(1)}(t)$ whose elements are exchanged from
eq. (\ref{eq:reduced_density_mat4}). Due to the fact that
$\mathcal{P}[\hat{\rho}_1^{(1)}(t)] =
\mathcal{P}[\hat{\rho}_2^{(1)}(t)] \leq 1$, the reduced density
matrices, $\hat{\rho}_1^{(1)}(t)$ and $\hat{\rho}_2^{(1)}(t)$,
represent mixed states, respectively, with both directly
reflecting the appearance of entanglement between the two coupled
oscillators; this is noted in
Fig.~\ref{fig:reduced_mat_time_evolution} with the properties of
$\mathcal{M}_{|\psi_1(t)\>}$ and $\mathcal{M}_{|\psi_2(t)\>}$ as
given in eq. (\ref{eq:entanglement_measure_case_1}). This
appearance of entanglement in the time evolution from the
disentangled initial product state for each case is clearly
attributed to the interaction Hamiltonian in
(\ref{eq:hamiltonian1}), i.e.,
$\hbar\,\kappa\,(\hat{a}_1^{\dagger}\,\hat{a}_2 +
\hat{a}_1\,\hat{a}_2^{\dagger})$. Likewise, for $J =
1,\frac{3}{2},5$ (or $N_1+N_2 = 2,3,10$), we also obtain the
entanglement in the time evolution resulting from the interaction
Hamiltonian. Due to their complex forms of eqs.
(\ref{eq:reduced_density_mat3_1}) and
(\ref{eq:entanglement_measure_case_1}) [see also
(\ref{eq:appendix_1}) and (\ref{eq:appendix_2})], respectively, we
simply plot the exact numerical results of
$\mathcal{M}_{|\psi(t)\>}$ for $N = N_1+N_2$ with $|\psi(0)\> =
|N\>|0\> = |J;J\>$ (see
Fig.~\ref{fig:reduced_mat_time_evolution}). We clearly see that
$\mathcal{M}_{|\psi(t)\>}$ increases with $N$ (or $J$).

Two points deserve comment here. First, we see that for a given
total occupation number $N = 2J$ [constant of motion; cf.
(\ref{eq:equation_of_motion1})], $\mathcal{M}_{|\psi(t)\>}$ is
periodic (see Fig.~\ref{fig:reduced_mat_time_evolution}), as is
the reduced density matrix $\hat{\rho}^{(1)}(t)$ due to the clear
oscillator dependence on time as seen in $f_{\mathbf{k}}^{(1)}$ in
(\ref{eq:f_function}); on the other hand, this is not the case for
a single oscillator coupled with a thermal reservoir modeled by a
sea of infinite oscillators \cite{LOU64}, for then, dissipative
decay of its occupation number into the the reservoir is
effective. Second, in calculating $\hat{\rho}^{(2)}(t) =
\mbox{Tr}_1\,\hat{\rho}(t)$ and comparing results to the
corresponding expressions in (\ref{eq:schroedinger_pic_psi2}) -
(\ref{eq:reduced_density_mat3}), one easily arrives at the fact
that $\hat{\rho}_{\mathbf{k}\mathbf{l}}^{(2)} =
\hat{\rho}_{-\mathbf{k},-\mathbf{l}}^{(1)}$ for any $J$, which
confirms the relationship valid for any system state $|\psi(t)\>$
that $\mathcal{P}[\hat{\rho}^{(2)}] =
\mathcal{P}[\hat{\rho}^{(1)}]$, thus indicating the equality of
mixture.

\begin{figure}[htbp]
    {\par\centering \resizebox*{0.45\textwidth}{!}
    {\includegraphics[angle=-90]{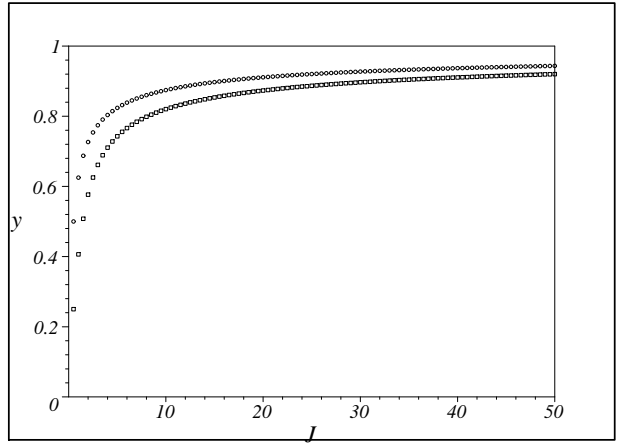}} \par}
    \caption{\label{fig:reduced_mat_time_evolution2}
    The measure of entanglement, $y = \mathcal{M}_{|\psi(t)\>}$, as given in
    eq. (\ref{eq:entanglement_measure_case_2}), versus
    $J$ is depicted for $\kappa = 1$ and
    initial states $|\psi(0)\> = |J;J\>_H = |N,0\>_H$ [cf.
    eq. (\ref{eq:eigen_initial_state_reduced_density_mat2})];
    the circles ($\circ$) are used to indicate $\gamma=\frac{\pi}{2}$ while the
    boxes ($\scriptscriptstyle{\Box}$) indicate $\gamma =
    \frac{\pi}{4}$. As noted in eq. (\ref{eq:case2_f}), $y$ is time-independent.}
\end{figure}

Next, we consider case II, in which the initial state is given by
an eigenstate of the exact Hamiltonian in eq.
(\ref{eq:coupled_boson_hamiltonian}), $\left|N_1,N_2\right>_{H} =
\left|J;M\right>_{H}$ with $J=\frac{1}{2}(N_1+N_2)$ and
$M=\frac{1}{2}(N_1-N_2)$. From eq.
(\ref{eq:reduced_density_mat2}), we easily obtain the reduced
density matrix and the measure of entanglement,
\begin{equation}\label{eq:eigen_initial_state_reduced_density_mat1}
    \hspace*{-3mm}{\textstyle
    \hat{\rho}_{\mathbf{k}\mathbf{l}}^{(1)}}\; =\;
    {\textstyle \delta_{\mathbf{k}\mathbf{l}} \cdot
    \left|\left<J;M \left|e^{i \gamma \hat{J}_y/\hbar}\right|J;\mathbf{k}\right>\right|^2\,
    =\, \delta_{\mathbf{k}\mathbf{l}} \cdot
    |f_{\mathbf{k}}^{(2)}|^2}\,,
\end{equation}
where
\begin{equation}\label{eq:case2_f}
    f_{\mathbf{k}}^{(2)}\; \equiv\;
    \hat{U}_{M\mathbf{k}}^{(J)}(\gamma)\,.
\end{equation}
Here, $f_{\mathbf{k}}^{(2)}$ is time-independent, whereas
$f_{\mathbf{k}}^{(1)}(t)$ from eq. (\ref{eq:f_function}) is not as
obtained in the previous case. Also, we can easily show that
$\hat{\rho}_{\mathbf{k}\mathbf{l}}^{(2)} =
\hat{\rho}_{-\mathbf{k},-\mathbf{l}}^{(1)}$. Then, as in eq.
(\ref{eq:entanglement_measure_case_1}), we have
\begin{equation}\label{eq:entanglement_measure_case_2}
    \hspace*{-7mm}{\textstyle \mathcal{M}_{|\psi(t)\>}}\; =\; 1\, -\, \sum_{\mathbf{k}=-J}^{J}\,
    {\textstyle |f_{\mathbf{k}}^{(2)}|^4\;.}
\end{equation}
Particularly for $M = J$, namely $N_2 = 0$, from eq.
(\ref{eq:rotation_mat2}) with $U_{J\mathbf{k}}^{(J)}(\gamma) =
U_{\mathbf{k}J}^{(J)}(-\gamma)$, we easily find that
\begin{equation}\label{eq:eigen_initial_state_reduced_density_mat2}
    {\textstyle f_{\mathbf{k}}^{(2)}\; =\;
    U_{J\mathbf{k}}^{(J)}(\gamma)\; =\;
    \sqrt{\frac{(2J)!}{k!\,(2J-k)!}}\,
    \left(\cos \frac{\gamma}{2}\right)^{k}\,
    \left(\sin \frac{\gamma}{2}\right)^{2J-k}\,,}
\end{equation}
where $k = J + \mathbf{k}$. Therefore, from eq.
(\ref{eq:eigen_initial_state_reduced_density_mat1}) we arrive at
the fact that the occupation number $k$ in oscillator $1$ is
described by the binomial distribution ${\mathcal{B}}(k;\,2J,p_1)$
with a trial probability $p_1 = (\cos \frac{\gamma}{2})^2$, which
remains unchanged in the time evolution. Along the same line, for
oscillator $2$, we also have
\begin{equation}\label{eq:eigen_initial_state_reduced_density_mat3}
    {\textstyle f_{\mathbf{k}}^{(2)}\; =\;
    U_{J\mathbf{k}}^{(J)}(\gamma)\; =\;
    \sqrt{\frac{(2J)!}{k!\,(2J-k)!}}\,
    \left(\sin \frac{\gamma}{2}\right)^{k}\,
    \left(\cos \frac{\gamma}{2}\right)^{2J-k}}
\end{equation}
and, from eq. (\ref{eq:eigen_initial_state_reduced_density_mat1}),
the binomial distribution ${\mathcal{B}}(k;\,2J,p_2)$ with $p_2 =
(\sin \frac{\gamma}{2})^2$ for the occupation number $k$ in
oscillator 2. These reduced density matrices, $\hat{\rho}^{(1)}$
and $\hat{\rho}^{(2)}$, represent mixed states for $\gamma \ne 0$,
respectively, due to the fact that
$\mathcal{P}[\hat{\rho}_1^{(1)}] = \mathcal{P}[\hat{\rho}_2^{(1)}]
\leq 1$, thus indicating the entanglement between the two linear
oscillators, $0 \leq \mathcal{M}_{|\psi(t)\>}$ obtained from eq.
(\ref{eq:entanglement_measure_case_2}). Furthermore, as in the
previous case, $\mathcal{M}_{|\psi(t)\>}$ increases with $J$ (see
Fig.~\ref{fig:reduced_mat_time_evolution2}); it is noted in the
figure that $\mathcal{M}_{|\psi(t)\>} \to 1$ as $J \to \infty$.
%
\section{Interaction between a linear and a non-linear
oscillator}\label{sec:linear_non_linear}
\subsection{Quantum-classical behaviors in the Heisenberg
picture}\label{subsec:linear_non_linear_heisenberg}
The second coupled system under investigation consists of a linear
oscillator and a (non-linear) angular momentum oscillator. This
system approximates the interaction between a field mode and an
atomic $N$-level system under idealized conditions resulting from
neglecting dissipation and the coupling between other surrounding
atomic systems. Following the seminal work of Ref.
\onlinecite{SEN71}, which considered the analysis
quantum-mechanically and also classically to compare one case with
the other systematically, it is convenient to employ, for the
linear oscillator, dimensionless coordinate $\mathtt{x}$ and
momentum $\mathtt{p}$, where $\mathtt{x} \equiv \sqrt{\frac{m
\omega}{\hbar}}\, {\mathit{x}}$ and $\mathtt{p} \equiv
\frac{1}{\sqrt{m \hbar \omega}}\, {\mathit{p}}$\,, satisfying
$\left[\mathtt{x}, \mathtt{p}\right]_{\mathsf{p}} = i$, and, for
the angular momentum oscillator, dimensionless angular momentum
variables ${\mathtt{j}}_x,\,{\mathtt{j}}_y,\,{\mathtt{j}}_z$ with
$\left[{\mathtt{j}}_r, {\mathtt{j}}_s\right]_{\mathsf{p}} =
i\,\epsilon_{rst}\,{\mathtt{j}}_t$, where the brackets $[\,,
\,]_{\mathsf{p}}$ represent the commutators for the
quantum-mechanical description, and $i$ times the Poisson brackets
for the classical description, respectively.

The Hamiltonians of the two individual oscillators are then
expressed as
\begin{equation}\label{eq:hamiltonian_of_two_osc}
    H_{1}\; =\; \frac{1}{2}\, \hbar\,\omega_1
    \left(\mathtt{x}_1^2\, +\, \mathtt{p}_1^2\right)\;,\;
    H_{2}\; =\;
    \hbar\,\omega_2\, \left(\mathtt{j}_{z}\right)_2\,,
\end{equation}
respectively, and the interaction Hamiltonian is given in the {\em
rotating wave approximation} by
\begin{equation}\label{eq:int_hamiltonian1}
    H_{12}\; =\; \hbar\, \kappa\,
    \left(a_1\, \left({\mathtt{j}}_{+}\right)_2\; +\; a_1^{\dagger}\,
    \left(\mathtt{j}_{-}\right)_2\right)\,,
\end{equation}
leading to the total Hamiltonian
\begin{equation}\label{eq:total_hamiltonian}
    H\; =\; H_{0}\; +\; H_{12}\;;\; H_0\; \equiv\; H_1\; +\; H_2\,.
\end{equation}
Here, we have
\begin{equation}\label{eq:non_hermitian_variables}
    \hspace*{-5mm}a_1 \equiv {\textstyle \frac{1}{\sqrt{2}} \left(\mathtt{x}_1 +
    i\,\mathtt{p}_1\right)}\;,\; a_1^{\dagger} \equiv {\textstyle \frac{1}{\sqrt{2}}
    \left(\mathtt{x}_1 - i\,\mathtt{p}_1\right)}\;,\;
    {\textstyle \left(\mathtt{j}_{+}\right)_2 \equiv
    \frac{1}{\sqrt{2}}\left\{\left(\mathtt{j}_x\right)_2 + i
    \left(\mathtt{j}_y\right)_2\right\}}\;,\;
    {\textstyle \left(\mathtt{j}_{-}\right)_2 \equiv
    \frac{1}{\sqrt{2}}\left\{\left(\mathtt{j}_x\right)_2 - i
    \left(\mathtt{j}_y\right)_2\right\}}
\end{equation}
(classically, Hermitian conjugation corresponds to complex
conjugation). These non-Hermitian variables and
$\left(\mathtt{j}_z\right)_2$ clearly obey
\begin{equation}\label{eq:commutator_rel1}
    {\textstyle \left[a_1,\,a_1^{\dagger}\right]_{\mathsf{p}}\; =\; 1\,,\;
    \left[\left(\mathtt{j}_{+}\right)_2,\,\left(\mathtt{j}_{-}\right)_2\right]_{\mathsf{p}}\;
    =\; \left(\mathtt{j}_z\right)_2\,,}\;
    \left[\left(\mathtt{j}_{+}\right)_2,\,\left(\mathtt{j}_z\right)_2\right]_{\mathsf{p}}\;
    =\; -\left(\mathtt{j}_{+}\right)_2\,,\;
    \left[\left(\mathtt{j}_{-}\right)_2,\,\left(\mathtt{j}_z\right)_2\right]_{\mathsf{p}}\;
    =\; \left(\mathtt{j}_{-}\right)_2
\end{equation}
with all other brackets vanishing. From now on, let us restrict
ourselves to the resonant case, $\omega_1 = \omega_2 \equiv
\omega$. In order to make later calculations simpler, we then
introduce the reduced variables\cite{SEN69} specified, with the
aid of the identity (\ref{eq:identity_1}) [for the classical
description, the commutators therein have to be obviously replaced
by $i$ times Poisson brackets] and the rules in
(\ref{eq:commutator_rel1}), by
\begin{equation}\label{eq:reduced_variables1}
    {\textstyle A_1\; \equiv\; e^{-i H_0\,t/\hbar}\; a_1\;
    e^{i H_0\,t/\hbar}\; =\; a_1\; e^{i \omega t}\;,}\;\;
    {\textstyle A_1^{\dagger}\; \equiv\; e^{-i H_0\,t/\hbar}\;
    a_1^{\dagger}\; e^{i H_0\,t/\hbar}\; =\;
    a_1^{\dagger}\; e^{-i \omega t}}
\end{equation}
and, similarly,
\begin{equation}\label{eq:reduced_variables2}
    \hspace*{-2mm}{\textstyle \left(J_{+}\right)_2\; =\;
    \left(\mathtt{j}_{+}\right)_2\, e^{-i \omega t}\;,\;
    \left(J_{-}\right)_2\; =\; \left(\mathtt{j}_{-}\right)_2\,
    e^{i \omega t}\;,\; \left(J_z\right)_2\; =\; \left(\mathtt{j}_z\right)_2\,,}
\end{equation}
respectively. These reduced variables are easily seen to satisfy
the same bracket relations as the corresponding unreduced
variables, namely,
\begin{equation}\label{eq:commutator_rel2}
    {\textstyle \left[A_1,\,A_1^{\dagger}\right]_{\mathsf{p}}\; =\; 1\;,\;
    \left[\left(J_{+}\right)_2,\,\left(J_{-}\right)_2\right]_{\mathsf{p}}\;
    =\; \left(J_z\right)_2\;,}\;
    \left[\left(J_{+}\right)_2,\,\left(J_z\right)_2\right]_{\mathsf{p}}\;
    =\; -\left(J_{+}\right)_2\;,\;
    \left[\left(J_{-}\right)_2,\,\left(J_z\right)_2\right]_{\mathsf{p}}\;
    =\; \left(J_{-}\right)_2\,.
\end{equation}
The interaction Hamiltonian in (\ref{eq:int_hamiltonian1}) is now
rewritten as
\begin{equation}\label{eq:int_hamiltonian2}
    {\textstyle H_{12}\; =\; \hbar\,\kappa\,
    \left\{A_1\, \left(J_{+}\right)_2\; +\; A_1^{\dagger}\,
    \left(J_{-}\right)_2\right\}\,,}
\end{equation}
and the equations of motion in terms of the reduced variables will
be given by
\begin{equation}\label{eq:eq_of_motion1}
    i\,\hbar\, \dot{O}(t)\; =\; \left[O(t),\,
    H_{12}\right]_{\mathsf{p}}\,,
\end{equation}
where $O$ stands for any of the above reduced variables.

From eqs. (\ref{eq:commutator_rel2}), (\ref{eq:eq_of_motion1}) it
follows that
\begin{eqnarray}\label{eq:eq_of_motion2}
    &\dot{A}_1\; =\; -i\, \kappa\, \left(J_{-}\right)_2\;,\;
    \dot{A}_1^{\dagger}\; =\; i\, \kappa\, \left(J_{+}\right)_2\,,&\n\\
    &\dot{\left(J_{+}\right)}_2\; =\; -i\, \kappa\,
    A_1^{\dagger}\, \left(J_z\right)_2\;,\;
    \dot{\left(J_{-}\right)}_2\; =\; i\, \kappa\, A_1\,
    \left(J_z\right)_2\;,\;
    \dot{\left(J_z\right)}_2\; =\; -i\, \kappa\,
    \left\{A_1\, \left(J_{+}\right)_2\, -\, A_1^{\dagger}\,
    \left(J_{-}\right)_2\right\}&
\end{eqnarray}
and then
\begin{equation}\label{eq:eq_of_motion3}
    \hspace*{-1cm}\ddot{\left(J_z\right)}_2\; =\; {\textstyle -\kappa^2\,
    \left\{\left(J_{-}\right)_2\,\left(J_{+}\right)_2\; +\;
    \left(J_{+}\right)_2\,\left(J_{-}\right)_2\; +\;
    \left(A_1\,A_1^{\dagger}\; +\;
    A_1^{\dagger}\,A_1\right)\,
    \left(J_z\right)_2\right\}}\;
    =\; {\textstyle -2\, \kappa^2\,
    \left\{\left(J_{-}\right)_2\,\left(J_{+}\right)_2\; +\;
    A_1\,A_1^{\dagger}\, \left(J_z\right)_2\right\}}\,,
\end{equation}
which, as shown previously,\cite{SEN71} hold both
quantum-mechanically in the Heisenberg picture and classically on
the basis of the Poisson bracket in phase space.

From the equation of motion in eq.~(\ref{eq:eq_of_motion1}) it
turns out that the sum of the (dimensionless) energy of both
oscillators, ${\mathtt{E}} = n_1 + \left(J_z\right)_2$ with $n_1
\equiv A_1^{\dagger}\,A_1$ (without the zero-point energy of the
linear oscillator in the quantum-mechanical description), and the
interaction Hamiltonian $H_{12}$ are constants of motion,
respectively, which are determined by a given initial state. As
well, the square of a given total angular momentum of the angular
momentum oscillator, $\mathbf{J}^2 = \left(J_x\right)_2^2 +
\left(J_y\right)_2^2 + \left(J_z\right)_2^2 = \left(J_{+}\right)_2
\left(J_{-}\right)_2 + \left(J_{-}\right)_2 \left(J_{+}\right)_2 +
\left(J_z\right)_2^2$\,, is also a constant of motion. Then,
substituting the expressions
\begin{equation}\label{eq:qm_commutator_rel}
    A_1\,A_1^{\dagger}\; =\; \mathtt{E}\, -\, \left(J_z\right)_2\, +\,
    \left[A_1,\, A_1^{\dagger}\right]\;,\;\;
    2\, \left(J_{-}\right)_2\,\left(J_{+}\right)_2\; =\; \mathbf{J}^2\, -\,
    \left(J_z\right)_2^2\, -\, \left[\left(J_{+}\right)_2,\,
    \left(J_{-}\right)_2\right]
\end{equation}
(note that there are no subscripts $\mathsf{p}$\, of the brackets)
into eq.~(\ref{eq:eq_of_motion3}), we obtain
\begin{equation}\label{eq:eq_of_motion4}
    \ddot{\left(J_z\right)}_2\; =\; \kappa^2
    \left\{3\,\left(J_z\right)_2^2\, -\,
    2 \left({\mathtt{E}}\, +\, \left[A_1,\, A_1^{\dagger}\right]\right)\,
    \left(J_z\right)_2\, +\,
    \left[\left(J_{+}\right)_2,\, \left(J_{-}\right)_2\right]\, -\,
    {\mathbf{J}}^2\right\}\,.
\end{equation}
Let us introduce the notation
\begin{equation}\label{eq:lambda1_2}
    \left[A_1,\, A_1^{\dagger}\right]\, =\, \lambda_1\;,\;\;
    \left[\left(J_{+}\right)_2,\, \left(J_{-}\right)_2\right]\, =\,
    \lambda_2\, \left(J_z\right)_2\,,
\end{equation}
where $\lambda_k=1,0$ with $k=1,2$ correspond to the
quantum-mechanical and the classical description of each
oscillator, respectively. Eq.~(\ref{eq:eq_of_motion4}) is now
reduced to
\begin{equation}\label{eq:eq_of_motion5}
    \ddot{\left(J_z\right)}_2\; =\; \kappa^2
    \left\{3\,\left(J_z\right)_2^2\, -\,
    \left(2\,\mathtt{E}\, +\, 2\,\lambda_1\, -\, \lambda_2\right)\,\left(J_z\right)_2\,
    -\, j\, \left(j\, +\, \lambda_2\right)\right\}\,,
\end{equation}
where the constant of motion ${\mathbf{J}}^2 \doteq j
\left(j+\lambda_2\right)$ with $j$ being classically
($\lambda_2=0$) the total (continuous) angular momentum and
quantum-mechanically ($\lambda_2=1$) the corresponding quantum
number. Here, $\left(J_z\right)_2$ is, clearly, an operator for
$\lambda_2 = 1$, with the property $\<\left(J_z\right)_2^2\> \ne
\<(J_z)_2\>^2$ in general. Eq. (\ref{eq:eq_of_motion5}) is a
non-linear differential equation for $\left(J_z\right)_2$ in both
the classical and quantum-mechanical description. Since
$[\left(J_z\right)_2, {\mathtt{E}}]_{\mathsf{p}} =
[\dot{\left(J_z\right)}_2, {\mathtt{E}}]_{\mathsf{p}} = 0$ (note
that ${\mathtt{E}} = n_1 + \left(J_z\right)_2$), it follows that
in the quantum-mechanical consideration of
eq.~(\ref{eq:eq_of_motion5}) the constant of motion ${\mathtt{E}}$
can be treated as a $c$-number.

We now consider eq.~(\ref{eq:eq_of_motion5}) with two interesting
initial states. First, for the ground state with $n_1=0$ and
$\left(J_z\right)_2=-j$, then $\mathtt{E} \equiv
{\mathtt{E}}_{g}=-j$; it immediately follows that eq.
(\ref{eq:eq_of_motion5}), at $t=0$, becomes
\begin{equation}\label{eq:eq_of_motion6}
    {\textstyle \left.\ddot{\left(J_z\right)}_2\right|_{t=0}}\; =\; 2\,\kappa^2\, \left(\lambda_1\, -\,
    \lambda_2\right)\, j\,.
\end{equation}
Quantum-mechanically, this state corresponds obviously to
$\left|\psi_g\right\> \equiv \left|0\right\>_1
\left|j,-j\right\>_2$. From the fact that each oscillator with the
minimum energy cannot give up further energy, we necessarily have
$\ddot{\left(J_z\right)}_2=0$ here, which indicates that only
$\lambda_1 \equiv \lambda_2 = 0,1$ simultaneously are physically
allowed. For the quantum description, the zero-point fluctuation
$\lambda_1=1$ would yield a positive force on the angular momentum
oscillator, and the fluctuation $\lambda_2=1$ of the angular
momentum oscillator would lead to energy transfer into the linear
oscillator; however, the two formal processes are always cancelled
such that they are not part of the real (or measurable) physical
processes of energy transfer.\cite{SEN95} Therefore, the
``semiclassical'' option with $\lambda_1=0$ and $\lambda_2=1$, and
vice versa, is not physically admissible for the ground state
$\left|\psi_g\right\>$, and is thus inappropriate for the system
analysis.

Another initial state under investigation is the state
${\mathtt{E}}=j$ with $n_1=0$ and $\left(J_z\right)_2=j$,
quantum-mechanically $\left|\psi_{\mathbf{J}}\right\> \equiv
\left|0\right\>_1 \left|j,j\right\>_2$. Then,
eq.~(\ref{eq:eq_of_motion5}), at $t=0$, reduces to
\begin{equation}\label{eq:eq_of_motion8}
    {\textstyle \left.\ddot{\left(J_z\right)}_2\right|_{t=0}}\; =\;
    -2\,\kappa^2\,\lambda_1\,j\,,
\end{equation}
independent of $\lambda_2$! This demonstrates that only for the
quantum description of the linear oscillator ($\lambda_1 = 1$) is
spontaneous emission available, which arises from the zero-point
fluctuation. Interestingly, it points out that for $\lambda_1 =
0$, there exists an unstable equilibrium leading to the absence of
spontaneous emission; the unstable equilibrium in the classical
description ($\lambda_1 = \lambda_2 = 0$) and the presence of
spontaneous emission for $\lambda_1 = 1$ were thoroughly discussed
in Ref. \onlinecite{SEN71}. Here, we connect the spontaneous
emission process with the notion of entanglement. In considering
the time evolution of the wave-function for the spontaneous
emission in the Schr\"{o}dinger picture, $\left|\psi(t)\right\> =
e^{-i H t/\hbar} \left|0\right\>_1 \left|j,j\right\>_2$ with the
Hamiltonian $H$ in eq. (\ref{eq:total_hamiltonian}), we now
necessarily have the entangled state for $t \to 0^{+}$, i.e.,
early time beyond the initial time, which is given by
\begin{equation}\label{eq:spontan_emission_entangled}
    \left|\psi(t)\right\>\; =\; {\textstyle \left|0\right\>_1
    \left|j,j\right\>_2\, -\, i\,t\, \left\{\omega
    \left(j + \frac{1}{2}\right) \left|0\right\>_1
    \left|j,j\right\>_2\, -\,
    \kappa \sqrt{j} \left|1\right\>_1 \left|j,j-1\right\>_2\right\}\,
    +\, O(t^2)\,.}
\end{equation}
This entanglement is explicitly shown in a compact form for the
two-level case $(j=\frac{1}{2})$ later in eq.
(\ref{eq:2_level_schroedinger1}) and in the more complex form for
the three-level case $(j=1)$ in (\ref{eq:3_level_density_mat}).
Therefore, for the quantum description $\lambda_1=\lambda_2=1$,
the entanglement is always temporally present in the spontaneous
emission process. Here, interestingly enough, we still have the
spontaneous emission for large enough $j$, which immediately leads
to the fact that the entanglement persists even as the {\em
coupled} angular momentum oscillator is taken to the limit of a
large number of levels, a limit which would go over to the
classical limit for an {\em uncoupled} angular momentum
oscillator. In essence, the state
$\left|\psi_{\mathbf{J}}\right\>$ of the uppermost excited state
of a $j$-level angular momentum oscillator is always coupled to
the vacuum state of the linear oscillator.

From eqs.~(\ref{eq:int_hamiltonian2}), (\ref{eq:eq_of_motion2}) we
also obtain
\begin{equation}\label{eq:differential_eq_for_J3_dot_square1}
    \dot{\left(J_z\right)}_2^2\, +\, K^2\; =\;
    2\,\kappa^2\,
    \left\{A_1\, A_1^{\dagger}\,
    \left(J_{+}\right)_2\, \left(J_{-}\right)_2\, +\,
    A_1^{\dagger}\, A_1\,
    \left(J_{-}\right)_2\, \left(J_{+}\right)_2\right\}\,,
\end{equation}
where $K \equiv H_{12}/\hbar$, and subsequently, with the aid of
eqs. (\ref{eq:qm_commutator_rel}), (\ref{eq:lambda1_2}), the
expression
\begin{equation}\label{eq:differential_eq_for_J3_dot_square2}
    \hspace{-1cm}{\textstyle \dot{\left(J_z\right)}_2^2}\; =\; {\textstyle 2\,\kappa^2\,
    \left[\left\{\left(J_z\right)_2\, -\, \left(\mathtt{E}\, +\,
    \frac{\lambda_1}{2}\right)\right\}\,
    \left\{\left(J_z\right)_2^2\, -\,
    \left(j\, (j\, +\, \lambda_2)\, -\,
    \frac{1}{2} \lambda_1\, \lambda_2\right)\right\}\right]\, +\,
    \left(\mathtt{E}\, +\, \frac{\lambda_1}{2}\right)
    \lambda_1\, \lambda_2\, \kappa^2\, -\, K^2\,.}
\end{equation}
In Ref.~\onlinecite{SEN71}, it was indicated that in the classical
description ($\lambda_1 = \lambda_2 = 0$)
eq.~(\ref{eq:differential_eq_for_J3_dot_square2}) has the form of
an equation for the vertical position $J_z$ of a classical
spherical pendulum, where the constant of motion $K$ represents an
angular momentum about the vertical axis through the center. It
turns out that for $n_1=0$ and $\left(J_z\right)_2=j$ we get $K =
0$ from (\ref{eq:int_hamiltonian2}) with $A_1 =
\sqrt{n_1}\,e^{i\,\theta_1},\, A_1^{\ast} =
\sqrt{n_1}\,e^{-i\,\theta_1}$ and $J_{\pm} = \sqrt{(\mathbf{J}^2 -
(J_z)^2)/2}\,e^{\pm i\,\theta_2}$ (note that in the quantum
description the constant of motion
$\left\<\psi_{\mathbf{J}}\right|K^2\left|\psi_{\mathbf{J}}\right\>
\ne 0$) and then $\dot{\left(J_z\right)}_2^2 = 0$ from
(\ref{eq:differential_eq_for_J3_dot_square2}). With
$\ddot{\left(J_z\right)}_2 = 0$ in eq.~(\ref{eq:eq_of_motion8}),
this reveals that the initial state $\left(n_1=0,\,
\left(J_z\right)_2=j\right)$ corresponds to an unstable
equilibrium of the pendulum.

Here, it is also worthwhile pointing out here that the
semiclassical treatment of Jaynes-Cummings \cite{JAY63} reproduces
this spontaneous emission for $j=\frac{1}{2}$ only, with the same
decay rate as that given by the quantum-mechanical description;
this is accomplished by adding a phenomenological damping term to
the classical equation of motion of the linear oscillator which
would result from the interaction with the instantaneous
expectation value of the dipole moment of the angular-momentum
oscillator. Therefore, the Jaynes-Cummings model just provides an
``artificial'' picture of the spontaneous emission without
describing the actual physical processes involved therein (e.g.,
entanglement), and also without offering a direct way for its
extension to more than the two-level case $(j \geq 1)$.

\subsection{Appearance of the entanglement in the Schr\"{o}dinger picture}
\label{subsec:linear_non_linear_schroedinger}
In order to study the appearance of the entanglement between the
linear and the angular momentum oscillators in the quantum
description, we discuss the state evolution of the total system in
the Schr\"{o}dinger picture given by
\begin{equation}\label{eq:hamiltonian_resonance_case2}
    {\textstyle \hat{H}\, =\, \hbar\, \omega\, (\hat{n}_1 +
    \frac{1}{2})\, +\, \hbar\, \omega\, \hat{J}_z}\, +\, \hbar\,
    \kappa\, (\hat{A}_1\,\hat{J}_{+}\, +\, \hat{A}_1^{\dagger}\,\hat{J}_{-})
\end{equation}
from eqs. (\ref{eq:hamiltonian_of_two_osc}) -
(\ref{eq:total_hamiltonian}) with the reduced variables in eqs.
(\ref{eq:reduced_variables1}) - (\ref{eq:reduced_variables2}) for
the resonant case, $\omega_1 = \omega_2 \equiv \omega$. We will
explicitly consider below the measure of entanglement for the
cases of $j=\frac{1}{2},1,\frac{3}{2}$ of the angular momentum
oscillator. To this end, as will be seen, the $(2j+1) \times
(2j+1)$ irreducible matrix representation for the total system is
employed, which is also systematically applicable for $j \geq 2$.
Here, we assume that the given total energy (a constant of motion)
is given by $\mathtt{E} \equiv n_1 + J_z = n_1 + j \geq j$.

\subsubsection{Case $j=\frac{1}{2}$}
We first consider the case of $j=\frac{1}{2}$, namely, a two-level
angular momentum oscillator interacting with a linear oscillator.
For the total energy $\mathtt{E} = n_1 + j = n_1 + \frac{1}{2}$,
the basis $\{|n_1+1, -\frac{1}{2}\>, |n_1, \frac{1}{2}\>\}$ allows
us to rewrite eq. (\ref{eq:hamiltonian_resonance_case2}) in
explicit operator form\cite{COM64} as
\begin{eqnarray}\label{eq:matrix_form_2_levels}
    \hat{H} &=& {\textstyle (n_1+1)\,\hbar\,\omega\,
    \left\{\,\left|n_1+1, -\frac{1}{2}\right\>\left\<n_1+1,
    -\frac{1}{2}\right|\; +\;
    \left|n_1, \frac{1}{2}\right\>\left\<n_1,
    \frac{1}{2}\right|\,\right\}\; +}\n\\
    && {\textstyle \frac{1}{\sqrt{2}}\, \hbar\,\kappa\,
    \sqrt{n_1+1}}\;
    {\textstyle \left\{\,\left|n_1+1, -\frac{1}{2}\right\>\left\<n_1,
    \frac{1}{2}\right|\,
    +\, \left|n_1, \frac{1}{2}\right\>\left\<n_1+1,
    -\frac{1}{2}\right|\,\right\}\,.}
\end{eqnarray}
This can be written in a $2 \times 2$ matrix form as
\begin{equation}\label{eq:hamiltonian_2_levels}
    {\textstyle \hat{H}\; =\; (n_1+1)\,\hbar\,\omega\,\id_2\, +\,
    \frac{1}{\sqrt{2}}\, \hbar\,\kappa\, \sqrt{n_1+1}\,
    \hat{\sigma}_x\,,}
\end{equation}
where $\hat{\sigma}_x = 2 \hat{J}_x$ denotes the Pauli matrix.
Using the identity,
\begin{equation}\label{eq:identities1}
    e^{i \alpha \hat{\sigma}_{y}}\, \hat{\sigma}_{x}\, e^{-i \alpha
    \hat{\sigma}_{y}}\;
    =\; (\cos 2\alpha)\, \hat{\sigma}_{x}\, +\, (\sin 2\alpha)\,
    \hat{\sigma}_{z}
\end{equation}
with $\alpha=\frac{\pi}{4}$, we arrive at the diagonalized
Hamiltonian for eq. (\ref{eq:hamiltonian_2_levels}) as
\begin{equation}\label{eq:2_level_1}
    \hspace*{-3mm}\hat{H}_d\; =\; {\textstyle (n_1+1)\,\hbar\,\omega\, \id_2\, +\,
    \frac{\hbar\,\kappa}{\sqrt{2}}\, \sqrt{n_1+1}\, (\sin
    2\alpha)\, \hat{\sigma}_z\,.}
\end{equation}
Accordingly, we acquire the energy eigenstates as
\begin{equation}\label{eq:eigenstates_linear_angular_momentum}
    \hspace*{-1.5cm}|\mathbf{1}_2\>\; =\;
    {\textstyle e^{-i \frac{\pi}{4} \hat{\sigma}_{y}}\,\left|n_1+1,
    -\frac{1}{2}\right\>\; =\; \frac{1}{\sqrt{2}}\,
    (|n_1+1, -\frac{1}{2}\>\, +\, |n_1, \frac{1}{2}\>)}\;,\;\;
    |\mathbf{2}_2\>\; =\;
    {\textstyle e^{-i \frac{\pi}{4} \hat{\sigma}_{y}}\,\left|n_1,
    \frac{1}{2}\right\>\, =\, \frac{1}{\sqrt{2}}\,
    (-|n_1+1, -\frac{1}{2}\>\, +\, |n_1, \frac{1}{2}\>)}
\end{equation}
with the corresponding eigenvalues, $E_1 = (n_1+1)\,\hbar\,\omega
- \frac{\hbar\,\kappa}{\sqrt{2}} \sqrt{n_1+1}$ and $E_2 =
(n_1+1)\,\hbar\,\omega + \frac{\hbar\,\kappa}{\sqrt{2}}
\sqrt{n_1+1}$, respectively.

\begin{figure}[htbp]
    {\par\centering \resizebox*{0.45\textwidth}{!}
    {\includegraphics{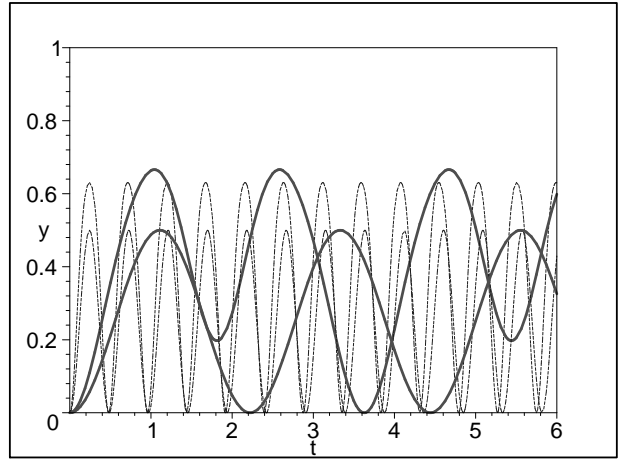}} \par}
    \caption{\label{fig:reduced_mat}
    The measure of entanglement, $y(t) = \mathcal{M}_{|\psi(t)\>}$ versus time
    $t$ is depicted for
    $\kappa = 1$, $\Delta \omega = 0$, and initial states $|\psi(0)\> =
    |n_1,j\>$; the solid lines are obtained for $n_1 = 0$, while the dash lines for $n_1 = 20$;
    $j = \frac{1}{2}, 1$ from bottom to top
    [cf. eqs. (\ref{eq:trace_square_reduced_density_mat}),
    (\ref{eq:3_level_density_mat})]; for $j =
    \frac{1}{2}$, the current $y(t)$ takes up the form of the $y(t)$ given in Fig.
    \ref{fig:reduced_mat_time_evolution}; for $j=1$, the current $y(t)$ becomes closer in form to the
    $y(t)$ given in Fig. \ref{fig:reduced_mat_time_evolution} with the increase of $n_1$.}
\end{figure}

Then, just as in discussion related to eq.
(\ref{eq:schroedinger_pic_psi1}) for the two coupled linear
oscillators, the state evolution for the initial state $|\psi(0)\>
= \left|n_1, \frac{1}{2}\right\>$ is given by
\begin{equation}\label{eq:2_level_schroedinger1}
    {\textstyle |\psi(t)\>\; =\; \frac{1}{\sqrt{2}}
    \left|\mathbf{1}_2\right\> e^{-\frac{i}{\hbar} E_1 t}
    - \frac{1}{\sqrt{2}} \left|\mathbf{2}_2\right\> e^{-\frac{i}{\hbar} E_2
    t}\,.}
\end{equation}
This state explicitly includes the possibility of spontaneous
emission ($n_1=0$). From eq. (\ref{eq:2_level_schroedinger1}), we
easily acquire the reduced density matrix of the angular momentum
oscillator as
\begin{equation}\label{eq:density_matrix_2_level_non_linear_osc}
    \hspace*{-5mm}\hat{\rho}^{(2)}(t)\; =\; \mbox{Tr}_1\,
    |\psi(t)\>\<\psi(t)|\;
    =\; {\textstyle \sin^2\left(\frac{\kappa t}{\sqrt{2}}
    \sqrt{n_1 + 1}\right)\,
    \left|-\frac{1}{2}\right\>\left\<-\frac{1}{2}\right|}\,
    +\, {\textstyle \cos^2\left(\frac{\kappa t}{\sqrt{2}} \sqrt{n_1 + 1}\right)\,
    \left|\frac{1}{2}\right\>\left\<\frac{1}{2}\right|\,,}
\end{equation}
which represents a mixed state. Therefore, it immediately follows
that
\begin{equation}\label{eq:trace_square_reduced_density_mat}
    \hspace*{-2mm}{\textstyle \mathcal{P}[\hat{\rho}^{(2)}]\; =\;
    \sin^4\left(\frac{\kappa t}{\sqrt{2}} \sqrt{n_1 + 1}\right)\,
    +\, \cos^4\left(\frac{\kappa t}{\sqrt{2}} \sqrt{n_1 + 1}\right)}\,,
\end{equation}
with $\mathcal{M}_{|\psi(t)\>} = 1 -
\mathcal{P}[\hat{\rho}^{(2)}]$ which oscillates between $0$ and
$\frac{1}{2}$ in the time evolution, thus indicating the
appearance of entanglement between the linear and the angular
momentum oscillators (see Fig.~\ref{fig:reduced_mat}).
Furthermore, from eq.
(\ref{eq:density_matrix_2_level_non_linear_osc}) we find, after a
minor calculation, that
\begin{equation}\label{eq:j_z_time_evolution}
    \hspace*{-1mm}{\textstyle \<\hat{J}_z\>}\; =\;
    {\textstyle \mbox{Tr}_2\,\{\hat{\rho}^{(2)}(t) \cdot
    \hat{J}_z\}\; =\; \frac{1}{2}\;,\;
    \cos \left(\frac{\kappa t}{\sqrt{2}} \sqrt{n_1+1}\right)\;,}\;
    {\textstyle \<\hat{J}_z^2\>}\;
    =\; {\textstyle \frac{1}{4}\;,\;
    \<\hat{J}_z^3\>\; =\; \frac{1}{8}\,
    \cos \left(\frac{\kappa t}{\sqrt{2}}
    \sqrt{n_1+1}\right)\,.}
\end{equation}
Therefore, from eqs. (\ref{eq:eq_of_motion5}),
(\ref{eq:j_z_time_evolution}), it follows that
\begin{equation}\label{eq:j_z_dot_square_time_evolution}
    {\textstyle \<\ddot{\hat{J}}_z\>\; =\; -2\,(n_1+1)\,
    \kappa^2\, \<\hat{J}_z\>\,,}
\end{equation}
which obviously yields the consistent result with that in eq.
(\ref{eq:eq_of_motion8}) for the case that $n_1 = 0$ and $J_z = j
= \frac{1}{2}$.

\subsubsection{Case $j=1$}
Next, we consider the case of $j=1$. For the total energy
$\mathtt{E} = n_1 + 1$, from eq.
(\ref{eq:hamiltonian_resonance_case2}) with the basis $\{|n_1+2,
-1\>,\, |n_1+1, 0\>,\, |n_1, 1\>\}$, the Hamiltonian takes up the
$3 \times 3$ irreducible matrix representation
\begin{equation}\label{eq:hamiltonian_3_level_matrix}
    \hat{H}\; \doteq\; \left(\begin{array}{ccc}
    {\textstyle \left(n_1+\frac{3}{2}\right)\, \hbar \omega} &
    {\textstyle \sqrt{n_1+2}\, \hbar \kappa} & {\textstyle
    0}\vspace*{2mm}\\
    {\textstyle \sqrt{n_1+2}\, \hbar \kappa} &
    {\textstyle \left(n_1+\frac{3}{2}\right)\, \hbar \omega} & {\textstyle
    \sqrt{n_1+1}\, \hbar \kappa}\vspace*{2mm}\\
    {\textstyle 0} & {\textstyle \sqrt{n_1+1}\, \hbar \kappa} &
    {\textstyle \left(n_1+\frac{3}{2}\right)\, \hbar
    \omega}\end{array}\right)\,.
\end{equation}
Based on the generating operators of the group
$SU(n)$,\cite{MAH98} where $n=2j+1$ (see Appendix
\ref{sec:appendix2}), the Hamiltonian in eq.
(\ref{eq:hamiltonian_3_level_matrix}) can be reexpressed as
\begin{equation}\label{eq:hamiltonian_3_level}
    \hat{H}\; =\; {\textstyle \hbar\,\omega\, \left(n_1+\frac{3}{2}\right)\,
    \id_3\, +\, \hat{H}_{12}\;,}\;\;
    {\textstyle \hat{H}_{12}}\; =\; {\textstyle \hbar\,\kappa\, \sqrt{n_1+2}\, \hat{u}_{12}\, +\,
    \hbar\,\kappa\, \sqrt{n_1+1}\, \hat{u}_{23}\,,}
\end{equation}
where $\hat{u}_{12}$ and $\hat{u}_{23}$ are generating operators
of $SU(n)$, and their matrix forms are explicitly given in eq.
(\ref{eq:su_n_algebra}).

Here, it is interesting to note that for a large $n_1 \gg 1$ such
that $\sqrt{n_1+1} \approx \sqrt{n_1+2}$, the Hamiltonian in eq.
(\ref{eq:hamiltonian_3_level_matrix}) can be transformed
approximately to
\begin{equation}\label{eq:approximate_hamiltonian}
    \hat{H}\; \simeq\; {\textstyle \left(n_1+\frac{3}{2}\right)\, \hbar\,\omega\, \id_3\,
    +\, \sqrt{2\,(n_1+2)}\, \hbar\,\kappa\, \hat{J}_x\,,}
\end{equation}
which formally corresponds to the exact Hamiltonian for
$j=\frac{1}{2}$ in eq. (\ref{eq:hamiltonian_2_levels}) with
$\hat{J}_x = \hat{\sigma}_x/2$. Thus, for a large $n_1$, we have
an approximate diagonalized representation $\hat{H}_d$ for $j=1$,
based on the previous result.

For the case of general $n_1$, proceeding from eq.
(\ref{eq:hamiltonian_3_level_matrix}), we easily obtain the energy
eigenstates with respect to the basis $\{|n_1+2, -1\>,\, |n_1+1,
0\>,\, |n_1, 1\>\}$ as
\begin{equation}\label{eq:3_eigenstates}
    \hspace*{-1cm}\left|\mathbf{1}_3\right\>\; \doteq\; \frac{1}{\sqrt{2\,(2n_1+3)}}\, \left(\begin{array}{r}
    \sqrt{n_1+2}\vspace*{2mm}\\
    \sqrt{2n_1+3}\vspace*{2mm}\\
    \sqrt{n_1+1}\end{array}\right)\;,\;
    \left|\mathbf{2}_3\right\>\; \doteq\; \frac{-1}{\sqrt{2n_1+3}}\, \left(\begin{array}{c}
    \sqrt{n_1+1}\vspace*{1mm}\\
    0\vspace*{1mm}\\
    -\sqrt{n_1+2}\end{array}\right)\;,\;
    \left|\mathbf{3}_3\right\>\; \doteq\; \frac{1}{\sqrt{2\,(2n_1+3)}}\, \left(\begin{array}{r}
    \sqrt{n_1+2}\vspace*{2mm}\\
    -\sqrt{2n_1+3}\vspace*{2mm}\\
    \sqrt{n_1+1}\end{array}\right)\,,
\end{equation}
and the corresponding energy eigenvalues,
\begin{equation}\label{eq:3_eigenvalues}
    E_1\; =\; {\textstyle \left(n_1+\frac{3}{2}\right)\, \hbar\,\omega\, +\,
    \kappa\,\hbar\, \sqrt{2n_1+3}}\;,\;
    E_2\; =\; {\textstyle \left(n_1+\frac{3}{2}\right)\,
    \hbar\,\omega}\;,\;
    E_3\; =\; {\textstyle \left(n_1+\frac{3}{2}\right)\, \hbar\,\omega\, -\,
    \kappa\,\hbar\, \sqrt{2n_1+3}\,,}
\end{equation}
respectively. Therefore, the diagonalized Hamiltonian is given by
\begin{equation}\label{eq:hamiltonian_3_level_diagonal}
    \hat{H}_d\; =\; {\textstyle \hbar\,\omega\, \left(n_1 + \frac{3}{2}\right)\,
    \id_3\;
    +\; \hbar\,\kappa\, \sqrt{2n_1 + 3}}\;
    {\textstyle \left(\, \left|\mathbf{1}_3\right\>\left\<\mathbf{1}_3\right|\;
    -\; \left|\mathbf{3}_3\right\>\left\<\mathbf{3}_3\right|\, \right)\,.}
\end{equation}

In order to diagonalize the Hamiltonian $\hat{H}$ in eq.
(\ref{eq:hamiltonian_resonance_case2}) for higher manifolds of
$j$, it is instructive to systematically analyze the above
diagonalization of $\hat{H}$ and confirm for $j=\frac{1}{2},1$. In
seeking the rotational operation that diagonalizes $\hat{H}$ in
eq. (\ref{eq:hamiltonian_3_level}) to achieve $\hat{H}_d$, it is
necessary to consider the unitary, Euler-like transformation
\begin{equation}\label{eq:diadonal_process3}
    \hspace*{-1mm}\hat{H}_d\, =\, e^{i \alpha_3\, \hat{v}_{13}}\, e^{i \alpha_2\, \hat{v}_{23}}\,
    e^{i \alpha_1\, \hat{v}_{12}}\, \hat{H}\, e^{-i \alpha_1\,
    \hat{v}_{12}}\, e^{-i \alpha_2\, \hat{v}_{23}}\, e^{-i \alpha_3\, \hat{v}_{13}}
\end{equation}
comprising three sequential, non-commuting rotations with respect
to $(\hat{v}_{12}, \hat{v}_{23}, \hat{v}_{13})$ defined in
(\ref{eq:su_n_algebra2}), with Euler angles $(\alpha_1, \alpha_2,
\alpha_3)$. Then, as previously accomplished for the case of
$j=\frac{1}{2}$ in eqs. (\ref{eq:hamiltonian_2_levels}) -
(\ref{eq:2_level_1}), here we similarly choose $(\alpha_1 =
-\frac{\pi}{2}; \cos \alpha_2 = \sqrt{n_1+1}/\sqrt{2n_1+3},\, \sin
\alpha_2 = -\sqrt{n_1+2}/\sqrt{2n_1+3}\,;\, \alpha_3 =
-\frac{\pi}{4})$ to obtain $\hat{H}_d$ in eq.
(\ref{eq:hamiltonian_3_level_diagonal}).

This clearly shows that the diagonalization of the Hamiltonian in
eq. (\ref{eq:hamiltonian_3_level}) cannot be described by a simple
rotation such as $\hat{R}^{(j=1)}(\alpha,\beta,\gamma) =
e^{-i\alpha \hat{J}_z}\, e^{-i\beta \hat{J}_y}\, e^{-i\gamma
\hat{J}_z}$ of an angular momentum oscillator,\cite{MAH98} whereas
for $j=\frac{1}{2}$, the rotation
$\hat{R}^{(j=\frac{1}{2})}(0,\frac{1}{2},0)$ was used in eq.
(\ref{eq:identities1}) to diagonalize $\hat{H}$ in eq.
(\ref{eq:hamiltonian_2_levels}). However, for the case of $j=1$
and large $n_1 \gg 1$ given in eq.
(\ref{eq:approximate_hamiltonian}), the diagonalization can be
approximately accomplished by
$\hat{R}^{(1)}\left(0,\frac{\pi}{2},0\right)$. Finally, it can be
shown by induction, for general $j$, that the diagonalization of
the Hamiltonian in eq. (\ref{eq:hamiltonian_resonance_case2}) is
generally characterized by $\frac{1}{2}(n^2-n)$ rotation angles
$\{\alpha_1, \alpha_2, \cdots, \alpha_{\frac{1}{2}(n^2-n)}\}$ with
respect to the set $\{\hat{v}_{jk}\}$ as given in eq.
(\ref{eq:su_n_generators}), where $n=2j+1$.

Let us now consider the state evolution with the initial condition
$|\psi(0)\> = |n_1,1\>$; just as in eq.
(\ref{eq:schroedinger_pic_psi1}), from eqs.
(\ref{eq:3_eigenstates}), (\ref{eq:3_eigenvalues}) we find that
\begin{equation}\label{eq:3_level_schroedinger1}
    {\textstyle |\psi(t)\>}\; =\; {\textstyle
    \sqrt{\frac{n_1 + 1}{2(2 n_1 + 3)}}\,
    \left|\mathbf{1}_3\right\>\, e^{-\frac{i}{\hbar} E_1 t}\; +}\;
    {\textstyle \sqrt{\frac{n_1+2}{2n_1+3}}\, \left|\mathbf{2}_3\right\>\, e^{-\frac{i}{\hbar} E_2
    t}\; +\; \sqrt{\frac{n_1+1}{2(2n_1+3)}}\, \left|\mathbf{3}_3\right\>\, e^{-\frac{i}{\hbar}
    E_3 t}\,.}
\end{equation}
From this state, we obtain, after some calculations, the reduced
density matrix of the angular momentum oscillator,
\begin{eqnarray}\label{eq:3_level_density_mat}
    \hspace*{-4.5mm}\hat{\rho}^{(2)}(t) &=& {\textstyle \mbox{Tr}_1
    |\psi(t)\>\<\psi(t)|\; =\; \frac{1}{(2 n_1 + 3)^2}\; \times}\;
    {\textstyle \left[4 (n_1 + 1) (n_1 + 2)\, \sin^4\left(\frac{\sqrt{2 n_1 + 3}}{2}\,\kappa
    t\right)\,|-1\>\<-1|\; +\right.}\n\\
    && {\textstyle (n_1 + 1) (2 n_1 + 3)\, \sin^2\left(\sqrt{2 n_1 + 3}\,\kappa
    t\right)\,\,|0\>\<0|\; +}\;
    {\textstyle \left.\left\{2 (n_1 + 1)\, \cos^2\left(\frac{\sqrt{2 n_1 + 3}}{2}\,\kappa
    t\right)\; +\; 1\right\}^2\,\,|1\>\<1|\, \right]\,.}
\end{eqnarray}
This reduced density matrix represents a mixed state with
$\mathcal{P}[\hat{\rho}^{(2)}] \leq 1$ and
$\mathcal{M}_{|\psi(t)\>} \geq 0$, directly reflecting the
entanglement of the total state $|\psi(t)\>$ [here, eq.
(\ref{eq:3_level_density_mat}) for $j=1$ is comparable to eq.
(\ref{eq:density_matrix_2_level_non_linear_osc}) for
$j=\frac{1}{2}$]. Furthermore, by noting that $|\psi(t +
{\mathcal{T})}\> \equiv e^{-i
\hat{H}_d\,{\mathcal{T}}/\hbar}\,|\psi(t)\> = |\psi(t)\>$ for a
given $t$ with the diagonalized form $\hat{H}_d$ in eq.
(\ref{eq:hamiltonian_3_level_diagonal}), we can easily show that
the state evolution in eq. (\ref{eq:3_level_schroedinger1})
displays periodicity with period ${\mathcal{T}} =
\frac{2\pi}{\sqrt{2n_1+3}\, \kappa}$ as noted in
Fig.~\ref{fig:reduced_mat}. This means that the time evolution
operator $\hat{U}(t) = e^{-i \hat{H} t/\hbar}$, expressed here
with respect to the basis $\{\id_3, \hat{\lambda}_p|\,p=1,2,
\cdots, 3^2-1\}$ given in Appendix~\ref{sec:appendix2}, as
\begin{equation}\label{eq:periodicity_3}
    \hat{U}(t)\; =\; \id_3\, +\, \sum _{p=1}^{3^2-1}
    U_p(t)\; \hat{\lambda}_p\,,
\end{equation}
where $U_p(t) \equiv \mbox{Tr}\,\{\hat{U}(t)\, \hat{\lambda}_p\}$
and $U_p(0)=0$ for all $p$, has the periodic property that $U_p(t)
= U_p(t+\mathcal{T})$.

\subsubsection{Case $j=\frac{3}{2}$}
Let us consider the case of $j=\frac{3}{2}$, where the Hamiltonian
of eq. (\ref{eq:hamiltonian_resonance_case2}) for the total energy
$\mathtt{E} = n_1 + \frac{3}{2}$ is given by
\begin{equation}\label{eq:hamiltonian_4_level}
    \hspace*{-5mm}\hat{H}\; =\; {\textstyle \hbar\,\omega\, \left(n_1 + 2\right)\,
    \id_4\, +\, \hat{H}_{12}\;;\;\; \hat{H}_{12}\; =\; \hbar\,\kappa}\;
    {\textstyle \left(\sqrt{\frac{3(n_1 + 3)}{2}}\, \hat{u}_{12}\, +\,
    \sqrt{2(n_1 + 2)}\, \hat{u}_{23}\, +\, \sqrt{\frac{3(n_1 + 1)}{2}}
\hat{u}_{34}\right)\,.}
\end{equation}
The diagonalized form of eq. (\ref{eq:hamiltonian_4_level}) can be
obtained, after tedious calculations similar to those described in
the previous case, as
\begin{equation}\label{eq:hamiltonian_4_level_diagonal}
    \hspace*{-3mm}\hat{H}_d\; =\; {\textstyle \hbar\,\omega\, \left(n_1 + 2\right)\, \id_4\,
    +\, \frac{\kappa\,\hbar}{2}\,\left\{\, \sqrt{\theta_1 + \theta_2}\,
    \left(\, \left|\mathbf{1}_4\right\>\left\<\mathbf{1}_4\right|\, -\,
    \left|\mathbf{2}_4\right\>\left\<\mathbf{2}_4\right|\,
    \right)\, +\, \sqrt{\theta_1 - \theta_2}\, \left(\,
    \left|\mathbf{3}_4\right\>\left\<\mathbf{3}_4\right|\, -\,
    \left|\mathbf{4}_4\right\>\left\<\mathbf{4}_4\right|\, \right)\, \right\}\,,}
\end{equation}
where $\theta_1 = 20 + 10\,n_1;\, \theta_2 = 2 \sqrt{73 + 64 n_1 +
16 n_1^2}$\,, and the $|\mathbf{p}_4\>$'s with $\mathbf{p} =
\mathbf{1}, \mathbf{2}, \mathbf{3}, \mathbf{4}$ denote energy
eigenvectors.

\begin{figure}[htbp]
    {\par\centering \resizebox*{0.45\textwidth}{!}
    {\includegraphics{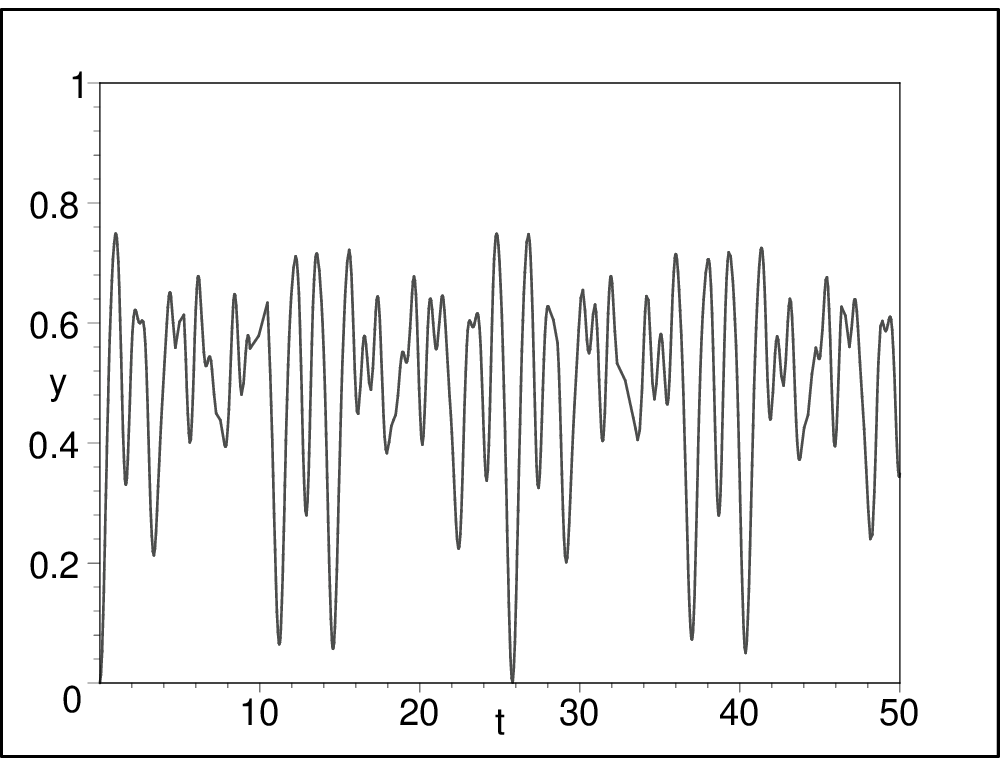}} \par}
    \caption{\label{fig:reduced_mat4_1}
    The measure of entanglement, $y(t) = \mathcal{M}_{|\psi(t)\>}$ versus time $t$
    is depicted for $\kappa = 1$,
    $\Delta \omega = 0$, and
    $|\psi(0)\> = |n_1=0,\,j=\frac{3}{2}\>$; we see here the the aperiodic behavior of $y(t)$, and
    its maximum $y_{\bf{m}}$ is larger than maxima of the $y(t)$ (solid lines) for $j = \frac{1}{2},
    1$ given in Fig. \ref{fig:reduced_mat}.}
\end{figure}

\begin{figure}[htbp]
    {\par\centering \resizebox*{0.45\textwidth}{!}
    {\includegraphics{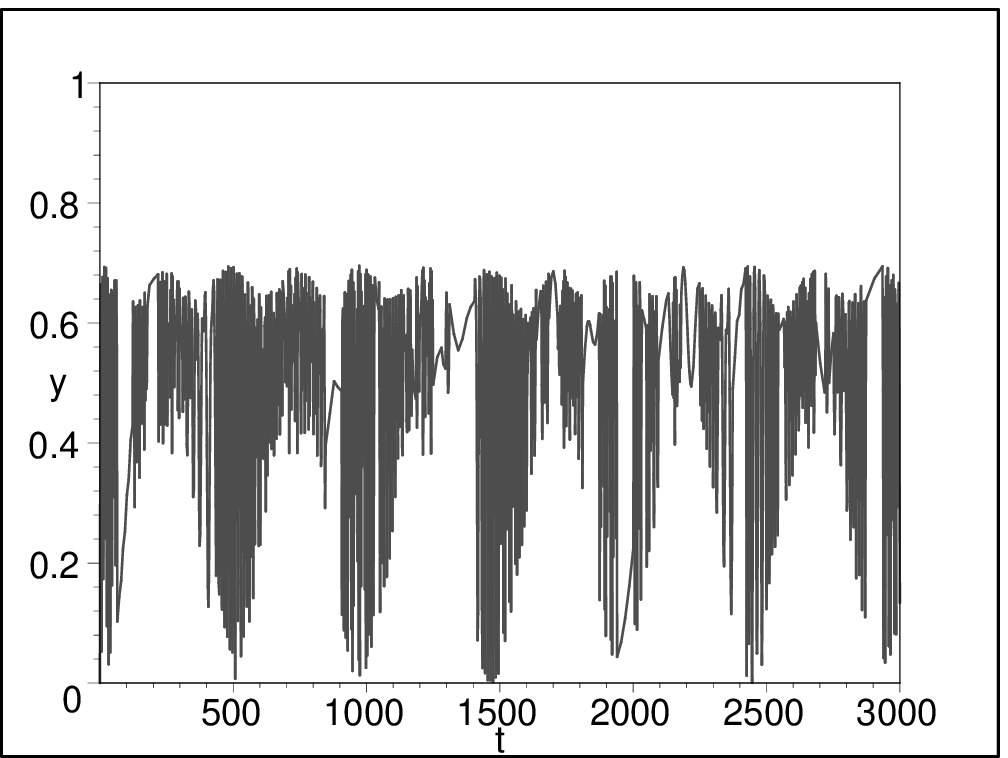}} \par}
    \caption{\label{fig:reduced_mat4_2}
    The measure of entanglement, $y(t) = \mathcal{M}_{|\psi(t)\>}$ versus time $t$
    is depicted for $\kappa = 1$, $\Delta \omega = 0$, and
    $|\psi(0)\> = |n_1=20,\,j=\frac{3}{2}\>$; we also here see the the aperiodic behavior of $y(t)$, and
    its maximum $y_{\bf{m}}$ is larger than maxima of the $y(t)$ (dash lines) for $j = \frac{1}{2}, 1$ given in
    Fig. \ref{fig:reduced_mat}.}
\end{figure}

Due to the complex form of the reduced density matrix of the
angular momentum oscillator, $\hat{\rho}^{(2)}(t) = \mbox{Tr}_1
|\psi(t)\>\<\psi(t)|$, and the measure of entanglement,
$\mathcal{M}_{|\psi(t)\>} = 1 - \mathcal{P}[\hat{\rho}^{(2)}(t)]$,
obtained from $|\psi(t)\> = e^{-i \hat{H} t/\hbar}\,|\psi(0)\>$
with $|\psi(0)\> = |n_1,\frac{3}{2}\>$, we simply plot the exact
numerical results of $\mathcal{M}_{|\psi(t)\>}$ in
Figs.~\ref{fig:reduced_mat4_1}, \ref{fig:reduced_mat4_2}.

Based on the numerical analyses\cite{HAK64} and the statistical
approximation in terms of classical random variables\cite{SEN71},
it has been shown that ${\textstyle \<\hat{J_z}(t)\>}$ with
$|\psi(0)\> = |0,j\>$ exhibits aperiodic behavior in its time
evolution for all $j \geq \frac{3}{2}$, even for $j$ large enough
(i.e., in the classical limit for an uncoupled angular momentum
oscillator), while the classical counterpart displays periodic
motion on the spherical pendulum [cf.
eq.~(\ref{eq:differential_eq_for_J3_dot_square2}) with $\lambda_1
= \lambda_2 = 0$ and $K = 0$; note that for this initial state,
the angular momentum $K$ about the vertical axis through the
center vanishes]. This shows that the aperiodicity in the time
evolution would be of non-classical origin. We indicate this
aperiodicity for any $n = 2j + 1 \geq 4$ here by noting the
impossibility of having the periodic property $\hat{U}(t) \equiv
e^{-i\hat{H}t/\hbar} = \id_n$ for arbitrary $t$; from eq.
(\ref{eq:hamiltonian_4_level_diagonal}), we find, after some
calculations, that
\begin{eqnarray}\label{eq:4_level_aperiodic_time_evolution_operator}
    \hat{U}(t) &=& e^{-\frac{i}{\hbar}\,\hat{H}_d\,t}\\
    &=& e^{-i\omega
    t(n_1+2)}\,
    (\,e^{-i\frac{\kappa}{2}\,\sqrt{\theta_1+\theta_2}\,t}\,|\mathbf{1}_4\>\<\mathbf{1}_4|\,
    +\, e^{i\frac{\kappa}{2}\,\sqrt{\theta_1+\theta_2}\,t/2}\,|\mathbf{2}_4\>\<\mathbf{2}_4|\,
    +\, e^{-i\frac{\kappa}{2}\,\sqrt{\theta_1-\theta_2}\,t}\,|\mathbf{3}_4\>\<\mathbf{3}_4|\,
    +\,
    e^{i\frac{\kappa}{2}\,\sqrt{\theta_1-\theta_2}\,t}\,|\mathbf{4}_4\>\<\mathbf{4}_4|\,)\,.\n
\end{eqnarray}
Since here the phase factors, $\kappa\,\sqrt{\theta_1+\theta_2}/2$
and $\kappa\,\sqrt{\theta_1-\theta_2}/2$, are incommensurable with
respect to each other, we immediately note that not all $U_k(t)$
in the expression
\begin{equation}\label{eq:aperiodicity_4}
    \hat{U}(t)\; =\; \id_4\, +\, \sum _{k=1}^{4^2-1}
    U_k(t)\; \hat{\lambda}_k
\end{equation}
can simultaneously vanish periodically, which leads to the
aperiodicity in the state evolution $|\psi(t)\> =
\hat{U}(t)\,|\psi(0)\>$ (see Figs.~\ref{fig:reduced_mat4_1},
\ref{fig:reduced_mat4_2}). Along the same line, it can be shown
that we have the diagonalized Hamiltonian for any $j$ in form
\begin{eqnarray}\label{eq:hamiltonian_n_level_diagonal}
    \hspace*{-9mm}\hat{H}_d &=&
    {\textstyle \hbar\,\omega\, \left(n_1 + \frac{1}{2} + j\right)\, \id_n\,
    +\, \kappa\,\hbar\, \left\{\beta_1\,
    \left(\, \left|\mathbf{1}_n\right\>\left\<\mathbf{1}_n\right|\, -\,
    \left|\mathbf{2}_n\right\>\left\<\mathbf{2}_n\right|\, \right)\, +\, \cdots\,
    +\right.}\n\\
    && {\textstyle \beta_m\, \left(\,\left|\mathbf{(2m-1)}_n\right\>\left\<\mathbf{(2m-1)}_n\right|\, -\,
    \left|\mathbf{(2m)}_n\right\>\left\<\mathbf{(2m)}_n\right|\,\right)\, +\,
    \cdots\, +}\,
    {\textstyle \left.\beta_{n/2}\, \left(\,
    \left|\mathbf{(n-1)}_n\right\>\left\<\mathbf{(n-1)}_n\right|\, -\,
    \left|\mathbf{n}_n\right\>\left\<\mathbf{n}_n\right|\, \right)\, \right\}\,,}
\end{eqnarray}
where $n = 2j+1$, and $\beta_m = \beta_m(n_1)$ with $m = 1,2,
\cdots, \frac{n}{2}$; $|\mathbf{p}_n\>$'s denote energy
eigenvectors. Here, each $\beta_{m_1}$ is, in general,
incommensurable with any other $\beta_{m_2}$, where $m_1 \ne m_2$,
and for the case that $n$ is odd, one of $\beta_m$'s always
vanishes. Therefore, we have
\begin{eqnarray}\label{eq:time_evolution_n_level}
    \hspace*{-1cm}\hat{U}(t) &=& e^{-\frac{i}{\hbar}\,\hat{H}_d\,t}\; =\; e^{-i\omega
    t (n_1 + \frac{1}{2} + j)}\,
    (\,e^{-i\kappa\,\beta_1\,t}\,|\mathbf{1}_n\>\<\mathbf{1}_n|\,
    +\, e^{i\kappa\,\beta_1\,t}\,|\mathbf{2}_n\>\<\mathbf{2}_n|\,
    +\, \cdots\, +\, e^{-i\kappa\,\beta_m\,t}\,|\mathbf{(2m-1)}_n\>\<\mathbf{(2m-1)}_n|\,
    +\n\\
    && e^{i\kappa\,\beta_m\,t}\,|\mathbf{(2m)}_n\>\<\mathbf{(2m)}_n|\, +\, \cdots\, +\,
    e^{-i\kappa\,\beta_{n/2}\,t}\,|\mathbf{(n-1)}_n\>\<\mathbf{(n-1)}_n|\,
    +\, e^{i\kappa\,\beta_{n/2}\,t}\,|\mathbf{n}_n\>\<\mathbf{n}_n|\,)\,,
\end{eqnarray}
thus showing that $\hat{U}(t) \ne \id_n$ for any $t$ unless all
$\beta_{m}$ are commensurable with each other. From this, we
easily find that the aperiodic behavior in the state evolution,
$|\psi(t)\> = \hat{U}(t)\,|\psi(0)\>$, will survive and even
increase with $n \geq 4$, i.e., $j \geq \frac{3}{2}$ [note that
for $j=1$, we obviously have $\beta_1$ only, which leads to the
periodicity in the state evolution: $\hat{U}(t) = \id_3$ for the
case that $\kappa\,\beta_1\,t = \pm 2\pi, \pm 4\pi, \pm 6\pi,
\cdots$].
%
\section{Conclusions}
In summary, we have investigated the fundamental dynamics of two
interacting oscillators: in one scenario, two linear oscillators,
and in the other scenario, a linear and a non-linear oscillator
have been considered. For the first scenario, based on the coupled
boson representation we compactly described the equation of motion
in the Heisenberg picture and also systematically studied the
quantum entanglement in the time evolution of the total
wave-function $|\psi(t)\>$ for various initial states in the
Schr\"{o}dinger picture. From this, we found that the appearance
of entanglement in the time evolution from the disentangled
initial simple product state is attributed to the interaction
between the two individual oscillators. Also, the measure of
entanglement increases with the total occupation number.

For the second scenario, quantum versus classical behaviors have
been studied, based on the Heisenberg equation developed in terms
of relevant reduced kinematics operator variables and
parameterized commutator relations. By setting the corresponding
commutator relations to one or zero, respectively, the Heisenberg
equations are shown to describe the full quantum or classical
motion of the interaction system, thus allowing us to discern the
differences between the fully quantum and fully classical
dynamical picture. In addition, for this second scenario, in the
fully quantum-mechanical description, we considered special
examples of $j = \frac{1}{2}, 1, \frac{3}{2}$ for the coupled
angular momentum state, demonstrating the explicit appearances of
entanglement. The entanglement increases with $j$ and so persists
even as $j \to \infty$, a limit which would go over to the
classical picture for an {\em uncoupled} angular momentum
oscillator. This entanglement occurs because the uppermost excited
state of the $j$-level angular momentum oscillator is always {\em
coupled} to the vacuum state of the linear oscillator, a purely
quantum coupling which manifestly gives rise to spontaneous
emission. We have also shown that the dynamics of this scenario
can be systematically described by the irreducible matrix
representation based on the generating operators of the group
$SU(n)$, while that of the first scenario can be given, based on
the coupled boson representation, simply by the rotations of an
angular momentum oscillator. For the coupled linear-angular
momentum oscillator, this system was shown to display periodicity
in the measure of entanglement for $j=\frac{1}{2}$ and $j=1$,
whereas for $j=\frac{3}{2}$ and beyond, the measure of
entanglement was shown to be aperiodic; this aperiodicity is
apparent from the form of the diagonalized multi-level Hamiltonian
and the resulting structure of the time evolution operator.
%
\section{Acknowledgments}
The authors acknowledge the support of the Office of Naval
Research and the National Science Foundation for this work.
%
\appendix\section{Mathematical supplements to eq. (\ref{eq:f_function})}
\label{sec:appendix1}
From eqs. (\ref{eq:rotation_mat3}) and (\ref{eq:f_function}) with
the relation that $U_{m\,\mathbf{k}}^{(J)}(\gamma) =
U_{\mathbf{k}\,m}^{(J)}(-\gamma)$, we find that
\begin{eqnarray}\label{eq:appendix_1}
    \hspace*{-5mm}f_{\mathbf{k}}^{(1)}(t) &=&
    \textstyle{(-1)^M\, \sqrt{\frac{(J+M)!\, (J-M)!}{(J+\mathbf{k})!\, (J-\mathbf{k})!}}\,
    \left(\sin \frac{\gamma}{2}\right)^{M-\mathbf{k}}\,
    \left(\cos \frac{\gamma}{2}\right)^{\mathbf{M}+\mathbf{k}}\;
    \times}\n\\
    && \sum_{m=-J}^{J}\, {\textstyle (-1)^m\, \left(\cos
    \frac{\gamma}{2}\right)^{2m} \cdot
    \left\{{\mathcal{P}}_{J-M}^{(M-m,M+m)}(\cos \gamma)\right\}\,
    \cdot}\;
    {\textstyle \left\{{\mathcal{P}}_{J-m}^{(m-\mathbf{k},m+\mathbf{k})}(\cos
    \gamma)\right\}\, \cdot\, e^{-i m \bar{\omega} t}}\,.
\end{eqnarray}
Particularly for the case that $\gamma = \frac{\pi}{2}$, namely
$\omega_1 = \omega_2$, eq. (\ref{eq:appendix_1}) reduces to
\begin{equation}\label{eq:appendix_2}
    \hspace*{-5mm}f_{\mathbf{k}}^{(1)}(t)\; =\;
    {\textstyle \frac{1}{4^{J}}\, \sqrt{\frac{(J+M)!\, (J-M)!}{(J+\mathbf{k})!\, (J-\mathbf{k})!}}}\,
    \sum_{m=-J}^{J}\, {\textstyle e^{-i m \bar{\omega} t}\;
    \cdot}\; \sum_{p=0}^{J-M}\, \sum_{q=0}^{J-m}\,
    {\textstyle (-1)^{p+q} \cdot {J-m \choose p}\, {J+m \choose J-M-p}\, {J-\mathbf{k} \choose q}\, {J+\mathbf{k} \choose
    J-m-q}\,,}
\end{equation}
where ${n \choose k} = \frac{n!}{k!\,(n-k)!}$.

\section{Generating operators of the group SU(n)}\label{sec:appendix2}
The set of the generating operators is given by \cite{MAH98}
\begin{equation}\label{eq:set_of_su_n_generators}
    \hspace*{-3mm}\pmb{\hat{\mathbf{\lambda}}}\; =\; \{\hat{\lambda}_p|\,p=1,2, \cdots, n^2-1\}\;
    =\; \{\hat{u}_{12}, \hat{u}_{13},
    \hat{u}_{23}, \cdots, \hat{v}_{12}, \hat{v}_{13}, \hat{v}_{23},
    \cdots, \hat{w}_{1}, \hat{w}_{2}, \cdots,
    \hat{w}_{n-1}\}\,,
\end{equation}
where $|\pmb{\hat{\mathbf{\lambda}}}| = n^2-1$, and
\begin{equation}\label{eq:su_n_generators}
    \hspace*{-3mm}\hat{u}_{kl}\; =\; \hat{P}_{kl} + \hat{P}_{lk}\;;\;\;
    \hat{v}_{kl}\; =\; i\, (\hat{P}_{kl} - \hat{P}_{lk})\;;\;\;
    \hat{w}_{m}\; =\;
    {\textstyle -\sqrt{\frac{2}{m(m+1)}}\;
    (\hat{P}_{11} + \hat{P}_{22} + \cdots + \hat{P}_{mm} - m
    \hat{P}_{m+1,m+1})\,.}
\end{equation}
Here, $1 \leq k < l \leq n$, and $|\{\hat{u}_{kl}\}| =
|\{\hat{v}_{kl}\}| = \frac{1}{2}(n^2-n)$, where $\{\hat{u}_{kl}\}$
and $\{\hat{v}_{kl}\}$ denote $\{\hat{u}_{12}, \hat{u}_{13},
\hat{u}_{23}, \cdots, \hat{u}_{n-1,n}\}$ and $\{\hat{v}_{12},
\hat{v}_{13}, \hat{v}_{23}, \cdots, \hat{v}_{n-1,n}\}$,
respectively; $1 \leq m \leq n-1$, and $|\{\hat{w}_{m}\}| = n-1$,
where $\{\hat{w}_{m}\}$ denotes $\{\hat{w}_{1}, \hat{w}_{2},
\cdots, \hat{w}_{n-1}\}$. $\hat{P}_{kl} = |k\>\<l|$ represents a
projection operator for $k = l$, and a transition operator for $k
\ne l$. For $n=2$ (i.e., $j=\frac{1}{2}$), the generators
$(\hat{u}_{12}, \hat{v}_{12}, \hat{w}_{1})$ exactly correspond to
$(\hat{\sigma}_x, \hat{\sigma}_y, \hat{\sigma}_z)$. For $n=3$
(i.e., $j=1$), three generators $\{\hat{u}_{12}, \hat{u}_{23},
\hat{u}_{13}\}$, for example, have the matrix representations,
\begin{equation}\label{eq:su_n_algebra}
    \hat{u}_{12}\; =\; \hat{P}_{12}\, +\, \hat{P}_{21}\;
    \doteq\;
    \left(\begin{array}{ccc}
    0 & 1 & 0\\
    1 & 0 & 0\\
    0 & 0 & 0\end{array}\right)\;,\;
    \hat{u}_{23}\; =\; \hat{P}_{23}\, +\, \hat{P}_{32}\;
    \doteq\;
    \left(\begin{array}{ccc}
    0 & 0 & 0\\
    0 & 0 & 1\\
    0 & 1 & 0\end{array}\right)\;,\;
    \hat{u}_{13}\; =\; \hat{P}_{13}\, +\, \hat{P}_{31}\;
    \doteq\;
    \left(\begin{array}{ccc}
    0 & 0 & 1\\
    0 & 0 & 0\\
    1 & 0 & 0\end{array}\right)\,,
\end{equation}
respectively. Similarly, for $3$ generators $\{\hat{v}_{12},
\hat{v}_{23}, \hat{v}_{13}\}$, we have
\begin{equation}\label{eq:su_n_algebra2}
    \hspace*{-.3cm}\hat{v}_{12} \doteq
    \left(\begin{array}{rrr}
    0 & i & 0\\
    -i & 0 & 0\\
    0 & 0 & 0\end{array}\right)\;,\;
    \hat{v}_{23} \doteq
    \left(\begin{array}{rrr}
    0 & 0 & 0\\
    0 & 0 & i\\
    0 & -i & 0\end{array}\right)\;,\;
    \hat{v}_{13} \doteq
    \left(\begin{array}{rrr}
    0 & 0 & i\\
    0 & 0 & 0\\
    -i & 0 & 0\end{array}\right)\,.
\end{equation}
The set $\{\id_n, \hat{\lambda}_p|\,p=1,2, \cdots, n^2-1\}$ can be
used as a well-defined basis of the space of $n \times n$
matrices.

\newpage
\vspace*{2cm}

\newpage
\vspace*{2cm}

\newpage
\end{document}